\documentclass[lettersize,hidelinks,journal]{IEEEtran}

\usepackage{amsmath,amssymb,amsfonts}
\usepackage{array}
\usepackage[caption=false,font=normalsize,labelfont=sf,textfont=sf]{subfig}
\usepackage{textcomp}
\usepackage{stfloats}
\usepackage{url}
\usepackage{verbatim}
\usepackage{graphicx}
\usepackage{cite}
\usepackage{xcolor}
\usepackage{orcidlink}
\usepackage[acronym,shortcuts]{glossaries}
\usepackage{bm}
\usepackage{relsize}
\usepackage{soul}
\usepackage{algorithm}
\usepackage[noend]{algpseudocode}
\usepackage{algcompatible}
\usepackage{mathtools}
\usepackage{bm}
\usepackage{lipsum}
\usepackage{mathtools}
\usepackage{lipsum}
\usepackage[bottom]{footmisc}
\usepackage{tabularx}

\def\BibTeX{{\rm B\kern-.05em{\sc i\kern-.025em b}\kern-.08em
T\kern-.1667em\lower.7ex\hbox{E}\kern-.125emX}}

\newcommand{\trans}[0]{^{\mathsf{T}}}

\newacronym{PCA}{PCA}{principal component analysis}
\newacronym{MDSM}{MDSM}{multi-domain sparse modulation}
\newacronym{P2P}{P2P}{point-to-point}
\newacronym{OTAC}{AirComp}{over-the-air computing}
\newacronym{TX}{TX}{transmitter}
\newacronym{RX}{RX}{receiver}
\newacronym{IoT}{IoT}{Internet-of-things}
\newacronym{AI/ML}{AI/ML}{artifitial intelligence/machine learning}
\newacronym{SDR}{SDR}{semi-definite relaxation}
\newacronym{EVD}{EVD}{eigenvalue decomposition}
\newacronym{GR}{GR}{Gaussian randomization}
\newacronym{SCA}{SCA}{successive convex approximation}
\newacronym{BnB}{BnB}{branch and bound}
\newacronym{QT}{QT}{quadratic transform}
\newacronym{RQ}{RQ}{Rayleigh quotient}
\newacronym{SOCP}{SOCP}{second-order cone programming}
\newacronym{CDF}{CDF}{cumulative distribution function}
\newacronym{UF}{UF}{uniform-forcing}
\newacronym{AP}{AP}{access point}
\newacronym{RSDR}{R-SDR}{regularized semi-definite relaxation}
\newacronym{R-SDR}{R-SDR}{regularized SDR}
\newacronym{flops}{flops}{floating point operations}
\newacronym{ED}{ED}{edge device}
\newacronym{SINR}{SINR}{signal to interference-plus-noise ratio}
\newacronym{SIC}{SIC}{successive interference cancellation}
\newacronym{CSI}{CSI}{channel state information}
\newacronym{LoS}{LoS}{line-of-sight}
\newacronym{NLoS}{NLoS}{non-LoS}
\newacronym{RPE}{RPE}{radar parameter estimation}
\newacronym{OTFS}{OTFS}{orthogonal time frequency space}
\newacronym{AFDM}{AFDM}{affine frequency division multiplexing}
\newacronym{CRLB}{CRLB}{Cram{\`e}r-Rao lower bound}
\newacronym{BCRLB}{BCRLB}{Bayesian Cram{\`e}r-Rao lower bound}
\newacronym{BBI}{BBI}{Bayesian bilinear inference}
\newacronym{AoA}{AoA}{angle-of-arrival}
\newacronym{SNR}{SNR}{signal-to-noise ratio}
\newacronym{ML}{ML}{maximum likelihood}
\newacronym{MIMO}{MIMO}{multiple-input multiple-output}
\newacronym{SIMO}{SIMO}{single-input multiple-output}
\newacronym{SISO}{SISO}{single-input single-output}
\newacronym{MUSIC}{MUSIC}{multiple signal classification}
\newacronym{MU}{MU}{multi-user}
\newacronym{ROOT-MUSIC}{ROOT-MUSIC}{ROOT multiple signal classification}
\newacronym{JCAS}{JCAS}{joint communication and sensing}
\newacronym{JCR}{JCR}{joint communications and radar}
\newacronym{ISAC}{ISAC}{integrated sensing and communications}
\newacronym{3D}{3D}{three-dimensional}
\newacronym{2D}{2D}{two-dimensional}
\newacronym{1D}{1D}{one-dimensional}
\newacronym{BF}{BF}{beamforming}
\newacronym{ROI}{ROI}{region of interest}
\newacronym{mmWave}{mmWave}{millimeter-wave}
\newacronym{MF}{MF}{matched-filter}
\newacronym{DD}{DD}{delay-Doppler}
\newacronym{SotA}{SotA}{state-of-the-art}
\newacronym{ULA}{ULA}{uniform linear array}
\newacronym{QAM}{QAM}{quadrature amplitude modulation}
\newacronym{ISFFT}{ISFFT}{inverse symplectic finite Fourier transform}
\newacronym{SFFT}{SFFT}{symplectic finite Fourier transform}
\newacronym{ISI}{ISI}{inter-symbol interference}
\newacronym{AWGN}{AWGN}{additive white Gaussian noise}
\newacronym{MSE}{MSE}{mean-squared-error}
\newacronym{LMMSE}{LMMSE}{linear minimum mean square error}
\newacronym{RMSE}{RMSE}{root mean square error}
\newacronym{ESPRIT}{ESPRIT}{estimation of signal parameters via rotational invariant techniques}
\newacronym{OFDM}{OFDM}{orthogonal frequency division multiplexing}
\newacronym{OCDM}{OCDM}{orthogonal chirp division multiplexing}
\newacronym{BS}{BS}{base station}
\newacronym{UE}{UE}{user equipment}
\newacronym{JCEDD}{JCEDD}{joint channel estimation and data detection}
\newacronym{PDA}{PDA}{probabilistic data association}
\newacronym{PMF}{PMF}{probability mass function}
\newacronym{PBiGaBP}{PBiGaBP}{parametric bilinear Gaussian belief propagation}
\newacronym{PBiGAMP}{PBiGAMP}{parametric bilinear generalized approximate message passing}
\newacronym{GaBP}{GaBP}{Gaussian belief propagation}
\newacronym{FT}{FT}{frequency-time}
\newacronym{DFT}{DFT}{discrete Fourier transform}
\newacronym{IDFT}{IDFT}{inverse discrete Fourier transform}
\newacronym{TD}{TD}{time domain}
\newacronym{wlg}{w.l.g.}{without loss of generality}
\newacronym{CP}{CP}{cyclic prefix}
\newacronym{DAF}{DAF}{discrete affine Fourier}
\newacronym{DAFT}{DAFT}{discrete affine Fourier transform}
\newacronym{IDAFT}{IDAFT}{inverse discrete affine Fourier transform}
\newacronym{CPP}{CPP}{\textit{chirp-periodic} prefix}
\newacronym{IDZT}{IDZT}{inverse discrete Zak transform}
\newacronym{DZT}{DZT}{discrete Zak transform}
\newacronym{P/S}{P/S}{parallel-to-serial}
\newacronym{S/P}{S/P}{serial-to-parallel}
\newacronym{SBL}{SBL}{sparse Bayesian learning}
\newacronym{MPA}{MPA}{message passing algorithms}
\newacronym{EM}{EM}{expectation maximization}
\newacronym{sIC}{soft IC}{soft interference cancellation}
\newacronym{soft RG}{soft RG}{soft replica generation}
\newacronym{BG}{BG}{belief generation}
\newacronym{SGA}{SGA}{scalar Gaussian approximation}
\newacronym{CLT}{CLT}{central limit theorem}
\newacronym{PDF}{PDF}{probability density function}
\newacronym{QPSK}{QPSK}{quadrature phase-shift keying}
\newacronym{ICI}{ICI}{inter-carrier interference}
\newacronym{BER}{BER}{bit error rate}
\newacronym{DoF}{DoF}{degrees-of-freedom}
\newacronym{VGA}{VGA}{vector Gaussian approximation}
\newacronym{FD}{FD}{full-duplex}
\newacronym{NMSE}{NMSE}{normalized mean square error}
\newacronym{KL}{KL}{Kullback-Leibler}
\newacronym{LASSO}{LASSO}{least absolute shrinkage and selection operator}
\newacronym{FP}{FP}{fractional programming}
\newacronym{CC}{CC}{communication-centric}
\newacronym{RC}{RC}{raised-cosine}
\newacronym{RRC}{RRC}{root raised-cosine}
\newacronym{6G}{6G}{sixth-generation}
\newacronym{V2X}{V2X}{vehicle-to-everything}
\newacronym{LEO}{LEO}{low-earth orbit}
\newacronym{I/O}{I/O}{input-output}
\newacronym{CE}{CE}{channel estimation}
\newacronym{ICC}{ICC}{integrated communication and computing}
\newacronym{ISCC}{ISCC}{integrated sensing, communications and computing}
\newacronym{PAM}{PAM}{pulse amplitude modulation}
\newacronym{iid}{i.i.d.}{independent and identically distributed}
\newacronym{FH}{FH}{frequency-hopping}
\newacronym{JRC}{JRC}{joint radar and communication}
\newacronym{DFRC}{DFRC}{dual-function radar communications}
\newacronym{AF}{AF}{ambiguity function}
\newacronym{PRI}{PRI}{pulse repetition interval}
\newacronym{PRF}{PRF}{pulse repetition frequency}
\newacronym{PPH}{PPH}{polynomial-phase hopping}
\newacronym{QSM}{QSM}{quadrature spatial modulation}
\newacronym{CPM}{CPM}{continuous phase modulation}
\newacronym{ECCM}{ECCM}{electronic counter-countermeasures}
\newacronym{RFA}{RFA}{random frequency agility}
\newacronym{RPA}{RPA}{random pulse repetition interval (PRI) agility}
\newacronym{RFPA}{RFPA}{random frequency and PRI agility}
\newacronym{ASK}{ASK}{amplitude shift keying}
\newacronym{PSK}{PSK}{phase shift keying}
\newacronym{RIP}{RIP}{restricted isometry property}
\newacronym{OMP}{OMP}{orthogonal matching pursuit}
\newacronym{CIR}{CIR}{channel impulse response}
\newacronym{TDD}{TDD}{time division duplex}
\newacronym{CRKG}{CRKG}{channel reciprocity-based key generation}
\newacronym{FCM}{FCM}{Fuzzy C-means}
\newacronym{PIRS}{PIRS}{polar code-based information scheme}
\newacronym{CRC}{CRC}{cyclic redundancy check}
\newacronym{BDR}{BDR}{bit disagreement rate}
\newacronym{SQ}{SQ}{scalar quantization}
\newacronym{VQ}{VQ}{vector quantization}
\newacronym{AI}{AI}{artificial intelligence}
\newacronym{V2V}{V2V}{vehicle to vehicle}
\newacronym{IM}{IM}{index modulation}
\newacronym{SM}{SM}{spatial modulation}
\newacronym{SIM}{SIM}{spatial index modulation}
\newacronym{PLS}{PLS}{physical layer security}
\newacronym{CU}{CU}{communication user}
\newacronym{QoS}{QoS}{quality-of-service}
\newacronym{MMSE}{MMSE}{minimum mean squared error}
\newacronym{CRB}{CRB}{Cramér–Rao bound}
\newacronym{DL}{DL}{deep learning}
\newacronym{RIS}{RIS}{reconfigurable intelligent surfaces}
\newacronym{THz}{THz}{terahertz}
\newacronym{JFI}{JFI}{Jain’s fairness index}
\newacronym{AN}{AN}{artificial noise}

\begin{document}

\title{Fairness-Aware Secure Integrated Sensing and Communications with Fractional Programming}

\author{
    Ali Khandan Boroujeni\textsuperscript{\orcidlink{0009-0003-5007-8535}},~\IEEEmembership{Graduate Student Member,~IEEE,} \\
    Kuranage Roche Rayan Ranasinghe\textsuperscript{\orcidlink{0000-0002-6834-8877}},~\IEEEmembership{Graduate Student Member,~IEEE,} \\
    Giuseppe Thadeu Freitas de Abreu\textsuperscript{\orcidlink{0000-0002-5018-8174}},~\IEEEmembership{Senior Member, IEEE,}
    Stefan Köpsell\textsuperscript{\orcidlink{0000-0002-0466-562X}},~\IEEEmembership{Senior Member, IEEE,} \\
    Ghazal Bagheri\textsuperscript{\orcidlink{0009-0006-2740-8235}},~\IEEEmembership{Graduate Student Member,~IEEE,} 
    and
    Rafael F. Schaefer\textsuperscript{\orcidlink{0000-0002-1702-9075}},~\IEEEmembership{Senior Member, IEEE}
    \vspace{-2ex}
    \thanks{Ali Khandan Boroujeni, Stefan Köpsell, and Rafael F. Schaefer are with the Barkhausen Institut and Technische Universit\"at Dresden, 01067 Dresden, Germany (emails: ali.khandanboroujeni@barkhauseninstitut.org; \{stefan.koepsell,rafael.schaefer\}@tu-dresden.de).}
    \thanks{Kuranage Roche Rayan Ranasinghe and Giuseppe Thadeu Freitas de Abreu are with the School of Computer Science and Engineering, Constructor University (previously Jacobs University Bremen), Campus Ring 1, 28759 Bremen, Germany (emails: \{kranasinghe,gabreu\}@constructor.university).}
    \thanks{Ghazal Bagheri is with Technische Universit\"at Dresden, 01187 Dresden, Germany (email: ghazal.bagheri@tu-dresden.de).}
}



\maketitle

\begin{abstract}

We propose a novel secure \ac{ISAC} system designed to serve multiple \acp{CU} and targets. 
To that end, we formulate an optimization problem that maximizes the secrecy rate under constraints balancing both communication and sensing requirements.
To enhance fairness among users, an entropy-regularized fairness metric is introduced within the problem framework.
We then propose a solution employing an accelerated \ac{QT} with a non-homogeneous bound to iteratively solve two subproblems, thereby effectively optimizing the overall objective.
This approach ensures robust security and fairness in resource allocation for \ac{ISAC} systems.
Finally, simulation results verify the performance gains in terms of average secrecy rate, average data rate, and beam gain.

\end{abstract}


\glsresetall

\IEEEpeerreviewmaketitle

\vspace{-2ex}
\section{Introduction}
\label{sec:introduction}

The evolution of wireless communication systems toward \ac{6G} and beyond has underscored the importance of \ac{ISAC} systems, which unify radar sensing and communication functionalities within a unified spectrum and hardware framework. 
This convergence addresses the increasing demand for spectral efficiency, cost-effective hardware, and support for emerging applications such as autonomous vehicles, smart cities, augmented reality, and massive \ac{IoT} deployments \cite{SuTWC2021, DongTGCN2023, ZhangIoTJ2024, WeiCOMMAG2022}.
By enabling simultaneous sensing and communication, \ac{ISAC} systems offer significant advantages, including enhanced resource utilization, reduced hardware complexity, and improved system scalability \cite{RenTCOM2023, PengTVT2025, WeiIoTJ2023}. 
However, the integration of these dual functionalities introduces multifaceted challenges, including security vulnerabilities, interference management, equitable resource allocation, and computational complexity in \ac{MU} environments \cite{LiIoTJ2025, ZhuTCOM2024, SandovalACCESS2023, ZhuoCOMMAG2024}.

Security remains a critical concern in \ac{ISAC} systems, particularly in scenarios involving malicious targets or eavesdroppers that threaten both sensing accuracy and communication confidentiality \cite{SuTWC2021, DeligiannisTAES2018, ChuTVT2023, MitevBOOK2024}.
The shared spectrum and hardware increase the attack surface, as adversaries could exploit sensing signals to infer sensitive information or disrupt communication links \cite{ LiIoTJ2025, ZhaoWCL2025, XuCSNDSP2024, boroujeni2025frequencyhoppingwaveformdesign}. 
To counter these threats, \ac{PLS} techniques have gained prominence, leveraging the physical properties of wireless channels to enhance security without relying on conventional cryptographic methods \cite{PengTVT2025, LiIoTJ2025, GunluJSAIT2023}. 
For example, \cite{DongTGCN2023} proposed joint beamforming designs for dual-functional \ac{MIMO} radar and communication systems to maximize secrecy rates while ensuring reliable sensing performance.
Similarly, \cite{DeligiannisTAES2018} developed secrecy rate optimization strategies for \ac{MIMO} communication radar systems, emphasizing the trade-off between security and performance. 
Other works, such as \cite{RenTCOM2023} and \cite{ChuTVT2023}, have explored robust beamforming techniques to mitigate interference from malicious targets, while \cite{XuCSNDSP2024} and \cite{boroujeni2025frequencyhoppingwaveformdesign} investigated secure waveform designs to enhance \ac{PLS} in \ac{ISAC} systems.
These studies highlight the need for robust and adaptive strategies to ensure security in adversarial environments.

In \ac{MU} \ac{ISAC} systems, ensuring fairness in resource allocation is paramount to preventing performance disparities, particularly when users have diverse \ac{QoS} requirements \cite{ZhuTCOM2024, SandovalACCESS2023, AgarwalTVT2025, ChouGLOBECOM2014}. 
Traditional optimization approaches, such as sum-rate maximization \cite{SandovalACCESS2023, ShtaiwiTCOM2023, SinghOJCOMS2025}, often prioritize aggregate system performance, which may favor users with stronger channel conditions, leading to inequitable resource distribution. 
To address this, fairness-aware frameworks have been proposed, drawing inspiration from resource allocation metrics in shared computer systems \cite{JainTECHREPORT1998, ZhuTCOM2024, AgarwalTVT2025}. 
The entropy-based fairness measure introduced by \cite{JainTECHREPORT1998} provides a quantitative framework to balance performance between users, which has been adapted for wireless systems \cite{DouIoTJ2024, AgarwalTVT2025, ChouGLOBECOM2014}. 
In the context of \ac{ISAC}, fairness is particularly challenging due to the conflicting objectives of sensing and communication, which compete for limited resources \cite{DouIoTJ2024, XiongTIT2023, LyuWCL2024, li2025fairnessvsequalityrsmabased}. 
Recent studies have proposed fairness-driven beamforming and resource allocation strategies to ensure equitable performance in \ac{MU} \ac{ISAC} systems \cite{DouIoTJ2024, AgarwalTVT2025}, while \cite{NguyenTVT2022} introduced proportional fairness metrics to balance sensing and communication performance.

The optimization of \ac{ISAC} systems involves addressing complex, non-convex problems driven by the interplay of sensing, communication, and security constraints \cite{II-ShenTSP2018, ShenJSAC2024, ShiTSP2011, KhanTWC2020}. 
Fractional programming has emerged as a powerful tool for tackling such problems, offering tractable solutions for power control, beamforming, and scheduling \cite{ShenTSP2018, II-ShenTSP2018, KhanTWC2020, ChenWCSP2024}. 
The Lagrange dual complex \ac{QT}, as explored in \cite{ShenJSAC2024}, provides an efficient framework for decomposing non-convex problems into iterative subproblems, enabling scalable solutions for large-scale systems \cite{ShiTSP2011, ChenWCSP2024, UchimuraTWC2025}. 
This approach has been successfully applied to \ac{MIMO} radar systems \cite{FuhrmannASILOMAR2004, FuhrmannTAES2008, LiuTSP2020, QiCOMML2022} and secure \ac{ISAC} designs \cite{RenTCOM2023, PengTVT2025, ChuTVT2023}. 
Additionally, techniques such as weighted \ac{MMSE} optimization \cite{ShiTSP2011}, alternating direction method of multipliers (ADMM) \cite{BoydBOOK2011}, and successive convex approximation (SCA) \cite{RazaviyaynBOOK2013} have been employed to address the computational complexity of \ac{ISAC} systems. 
Recent advances, such as \cite{LuIoTJ2024} and \cite{LiuTWC2025}, have proposed hybrid optimization frameworks that combine fractional programming and machine learning to enhance convergence rates and adaptability.

\Ac{SotA} advancements in \ac{ISAC} systems have focused on improving sensing accuracy, communication reliability, security, and energy efficiency. 
For instance, \cite{LiuTSP2020} proposed joint transmit beamforming for multi-user \ac{MIMO} communications and radar, achieving enhanced target detection while maintaining communication quality. 
Similarly, \cite{LyuWCL2024} developed dual-functional beamforming strategies that optimize both \ac{SNR} for communication and \ac{CRB} for sensing, addressing the inherent trade-offs in \ac{ISAC} systems \cite{XiongTIT2023, HuaTWC2024, WangTWC2024}. 
Robust beamforming techniques have been proposed to mitigate interference from malicious targets \cite{ChuTVT2023, PengTVT2025, WangTSP2024, XuCSNDSP2024}, while \cite{MengWCSP2022} introduced adaptive waveform designs to enhance sensing performance in cluttered environments. 
Machine learning techniques have also gained traction, with \cite{LongIoTJ2024} proposing deep reinforcement learning for \ac{ISAC}, \cite{LiuJSAC2022} developing neural network-based beamforming strategies, and \cite{ZhangTVT2024} exploring \ac{DL} for joint sensing and communication. 
Energy efficiency has been addressed in works such as \cite{HuangWCL2023}, \cite{LiuICCT2023}, which proposed energy-efficient designs for \ac{ISAC} systems to meet the power constraints of practical deployments. 

Practical implementation of \ac{ISAC} systems faces several challenges, including hardware limitations, channel estimation errors, dynamic interference environments, and real-time computational complexity \cite{WeiCOMMAG2022, XuICC2023, ZhuoCOMMAG2024}. 
Accurate \ac{CSI} is critical for effective beamforming and resource allocation, yet imperfect \ac{CSI} can significantly degrade performance \cite{XuJSAS2024, XiaoTWC2025, ZhangGLOBECOM2023}.
To address this, \cite{XuICC2023} and \cite{XuJSAS2024} proposed robust channel estimation techniques for \ac{ISAC} systems, while \cite{XiaoTWC2025} investigated the impact of \ac{CSI} errors on system performance. 
Computational complexity remains a significant hurdle, particularly for large-scale systems with multiple users and targets \cite{ShenJSAC2024, LiaoGLOBECOM2023}. 
Low-complexity algorithms, such as those proposed by \cite{LiaoGLOBECOM2023}, \cite{TemizOJCOMS2025}, and \cite{ZargariWCL2025}, aim to reduce computational overhead while maintaining performance. 
Emerging technologies, such as \ac{RIS} \cite{ChepuriSPM2023, IsmailOJCOMS2024, SaikiaGLOBECOM2023} and \ac{THz} communications \cite{ElbirAESM2024, WuIRMMW2021}, offer new opportunities to enhance \ac{ISAC} performance by improving signal propagation and spectral efficiency. 
For instance, \cite{IsmailOJCOMS2024} explored \ac{RIS}-assisted \ac{ISAC} systems to enhance coverage and security, while \cite{ElbirAESM2024} investigated \ac{THz} bands for high-resolution sensing and ultra-high-speed communication. 
Future research directions include the integration of \ac{ISAC} with edge computing \cite{HaqueWiMob2024}, federated learning \cite{ZhangINFOCOM2024}, and quantum communication \cite{TariqIoTJ2024}, which will advance the capabilities of \ac{ISAC} systems.

In this paper, we propose a comprehensive and practical framework for secure multi-user \ac{ISAC} systems involving multiple \acp{CU} and multiple targets. 
This approach builds on prior works in \ac{MIMO} radar beamforming \cite{FuhrmannASILOMAR2004, FuhrmannTAES2008, LiuTSP2020, QiCOMML2022}, secure \ac{ISAC} designs \cite{RenTCOM2023, ChuTVT2023, PengTVT2025}, and fractional programming \cite{ShenTSP2018, II-ShenTSP2018, KhanTWC2020, ChenWCSP2024}. 
Unlike most existing works that consider only a single target, often under simplified assumptions that limit practical relevance, our formulation accounts for multiple targets, each with the potential to act as an eavesdropper. 
This introduces new challenges in simultaneously ensuring reliable communication, accurate sensing, and strong protection of confidential information in adversarial environments.
We aim to maximize the overall secrecy rate under strict communication and sensing constraints. A key limitation in prior works is the reliance on predefined SINR thresholds for each user, which leads to suboptimal and unfair resource allocation.
In particular, beamforming vectors are often optimized just to satisfy these minimum thresholds, favoring users with strong channels while neglecting others. This imbalance compromises fairness and degrades system utility in multi-user scenarios.

To overcome this issue, we introduce an entropy-regularized fairness metric, inspired by \ac{JFI}, which provides a continuous and interpretable measure of fairness in resource allocation. 
\ac{JFI} is particularly well-suited for wireless communications due to its scale invariance, bounded range, and computational simplicity, making it ideal for multi-objective optimization. 
Integrating \ac{JFI} into the beamforming design promotes equitable treatment of users while maintaining strong secrecy performance.
To further explore the tradeoff between fairness and throughput, we develop a novel iterative entropy-based tradeoff interpolation scheme that traverses a discretized fairness-throughput spectrum. 
This procedure enables smooth transitions between fairness-dominated and throughput-dominated regimes, providing deeper insights into system behavior and helping to identify balanced solutions tailored to application needs.
We solve the proposed non-convex optimization problem using an efficient Lagrangian dual framework combined with a non-homogeneous complex \ac{QT}, a novel solution for \ac{FP}. 
This approach decomposes the problem into two tractable subproblems, enabling joint optimization of secure beamforming and \ac{AN} design. 
Unlike traditional uses of \ac{AN}, which focus solely on jamming eavesdroppers, our method also leverages \ac{AN} to enhance sensing accuracy while aligning it within the null space of legitimate users, thereby avoiding interference with communication signals.
Finally, we demonstrate the practical benefits of our approach in large-scale systems, where computational efficiency and scalability are paramount. Our results confirm that the proposed framework is not only secure and fair, but also highly adaptable to emerging wireless technologies such as \ac{6G} \ac{ISAC} networks.

The remainder of this paper is organized as follows. Section \ref{sec:sys_model} presents the system model and problem formulation, including the proposed metrics to be optimized. 
Section \ref{sec:structuring_opt} details the full formulation of the optimization problem.
Section \ref{sec:dual_opt_frmework} presents the solution methodology based on the Lagrange dual complex \ac{QT}. 
Section \ref{sec:performance_analysis} provides simulation results and performance analysis to validate the proposed approach, and Section \ref{sec:conclusion} concludes the paper with future research directions.

The contributions of this paper are summarized as follows:
\begin{itemize}
\item We formulate a secure MU-ISAC system that considers multiple targets as potential eavesdroppers, capturing realistic adversarial settings.
\item A fairness-aware beamforming design is proposed by integrating an entropy-regularized Jain’s index, overcoming the limitations of fixed SINR thresholds.
\item An iterative tradeoff scheme is introduced to smoothly navigate the fairness–throughput spectrum, enabling flexible system tuning.
\item The non-convex problem is efficiently solved using a dual fractional programming approach that employs a non-homogeneous \ac{QT}, enabling closed-form updates that avoid costly matrix inversions and accelerate convergence.
\item \ac{AN} is jointly optimized to enhance both secrecy and sensing, while being confined to the null space of communication users to prevent interference.
\item The overall framework is scalable and well-suited for large-scale \ac{MIMO} \ac{ISAC} deployments in next-generation wireless networks.

\end{itemize}

\vspace{-2ex}
\section{System Model}

\label{sec:sys_model}
\begin{figure}[b!]
    \centering
    \includegraphics[width=\linewidth]{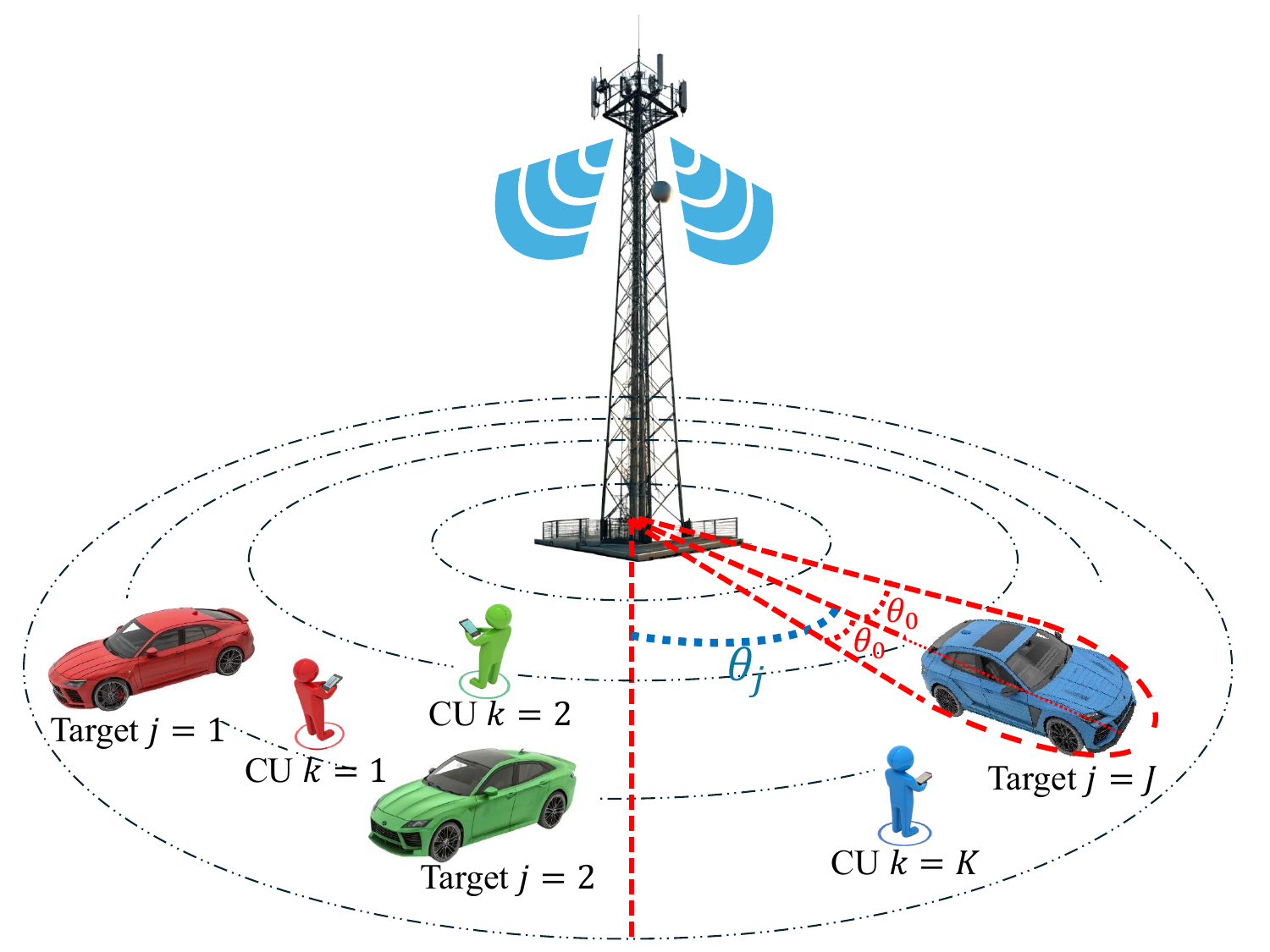}
    \caption{System model of a secure \ac{MU}-\ac{ISAC} network, where a multi-antenna \ac{BS} serves $K$ \acp{CU} while sensing $J$ targets that may act as eavesdroppers. Each target’s direction $\theta_j$ is assumed to be pre-estimated.}
    \label{fig:system_model}
\end{figure}

Consider a \ac{MU} environment with a single \ac{BS} employing countermeasures against potential eavesdroppers, hereafter referred to as Eves, which are represented by radar targets, while serving \acp{CU}.
As depicted in Fig. \ref{fig:system_model}, the \ac{BS} is equipped with \( N_t \) transmit antennas, enabling the use of multi-antenna signal processing techniques, such as beamforming and \ac{AN} injection, to both enhance communication performance and enable security features. 

In this scenario, the \ac{BS} serves as the central node for managing communication with \( K \) \acp{CU}, while localizing \( J \) radar targets (Eves), and ensuring secure information transmission by mitigating eavesdropping threats from the aforementioned targets.
The \( K \) legitimate \acp{CU}, each with a single receive antenna, are the intended recipients of the transmitted signals and leverage the spatial diversity and secure transmission strategies implemented by the \ac{BS} to enhance their communication performance.
The network also considers that the \( J \) untrusted targets are equipped with eavesdropping capabilities. 
The location of each radar target is assumed to be pre-estimated, with $\theta_j$ denoting the direction of the $j$-th target.
Ensuring secure communication against potential eavesdroppers is a critical aspect of system design, which we carefully address in the upcoming sections.

\vspace{-1ex}
\subsection{Transmit Signal Model}

Leveraging both transmit beamforming and \ac{AN}, the effective transmit signal can be defined as
\begin{equation}
\label{eq.transmit_signal}
    \boldsymbol{x} = \boldsymbol{W}\boldsymbol{s} + \boldsymbol{n_{\text{eff}}} \in \mathbb{C}^{N_t \times 1},
\end{equation}  
where \( \boldsymbol{s} \triangleq [s_1, \cdots, s_k, \cdots, s_K]\trans \in \mathbb{C}^{K \times 1} \) represents the symbol vector containing data intended for the \( K \) legitimate users. 
Here, \( s_k \) represents the input data symbols drawn from a \ac{PSK} constellation, where the uniform amplitude simplifies symbol normalization and enhances secrecy rate optimization. 
The beamforming matrix \( \boldsymbol{W} = [\boldsymbol{w}_1, \cdots, \boldsymbol{w}_k, \cdots, \boldsymbol{w}_K] \in \mathbb{C}^{N_t \times K} \), with $\boldsymbol{w}_k \in \mathbb{C}^{N_t \times 1}$ denoting the beamforming vector of the $k$-th user, is designed to optimize the secrecy rate for the intended \acp{CU} while reducing the risk of interception by potential eavesdroppers.
Notably, \( \boldsymbol{W} \) exploits the spatial diversity provided by the \( N_t \) antennas of the base station (BS).
The term $\boldsymbol{n}_{\text{eff}} \in \mathbb{C}^{N_t \times 1} \sim \mathcal{CN}(0, \boldsymbol{R}_{n_{\text{eff}}})$ denotes the effective artificial noise (AN) vector with covariance matrix \(\boldsymbol{R}_{n_{\text{eff}}}\), which is intentionally injected to degrade the interception capabilities of untrusted targets, enhance sensing performance, and create minimal interference to legitimate CUs while inducing significant disruption at potential eavesdroppers.

The primary objective of the system is to achieve secure and reliable communication for the \( K \) legitimate \acp{CU} while ensuring that the information leakage to the \( J \) untrusted targets is minimized. 
This is accomplished through the joint design of the beamforming matrix \( \boldsymbol{W} \) and the effective \ac{AN} \( \boldsymbol{n_{\text{eff}}} \), leveraging the multiple antennas at the BS and the advanced optimization techniques.

\vspace{-1ex}
\subsection{Received Signal Model}
Assuming that \acp{CU} can be treated as radar targets, and defining $Q\triangleq K + J$, the reflected echoes from these targets at the \ac{BS} can be expressed as
\begin{align}
    \boldsymbol{r} &= \underbrace{\sum\limits_{q=1}^{Q} \alpha_q^2 \boldsymbol{a}(\boldsymbol{\theta}_q)^{\text{T}} \boldsymbol{a}(\boldsymbol{\theta}_q)}_{\boldsymbol{G} \in \mathbb{C}^{N_t \times N_t}} \boldsymbol{x} + \boldsymbol{z} \nonumber \\
    &= \boldsymbol{G}(\boldsymbol{W}\boldsymbol{s} + \boldsymbol{n_{\text{eff}}}) + \boldsymbol{z}. \label{eq.received_signal} 
\end{align}
Let $\alpha_q$ denote the complex path gain from the transmitter to the potential target $q$. The transmit array steering vector ${\boldsymbol{a}}(\theta_q) \in \mathbb{C}^{N_t \times 1}$, corresponding to an arbitrary spatial direction $\theta_q$ in the far field, is given by
\begin{equation}
\label{eq:steering_vector}
{\boldsymbol{a}}(\theta_q) = \left[ 1,\ e^{j2\pi \frac{d_M}{\lambda} \sin(\theta_q)},\ \ldots,\ e^{j2\pi \frac{d_M}{\lambda}(N_T - 1)\sin(\theta_q)} \right]^{\text{T}},
\end{equation}
where $d_M$ represents the inter-element spacing of the transmit array (in meters), and $\lambda$ is the signal wavelength. The background \ac{AWGN} at the \ac{BS} is modeled as $\boldsymbol{z} \sim \mathcal{CN}(0, \sigma_r^2\boldsymbol{I})$, where $\sigma_r^2$ denotes the background noise power.

\subsection{Performance Metrics}
This paper aims to maximize the \textit{sum secrecy rate} of a \ac{MU} \ac{ISAC} system in the presence of multiple potential eavesdropping targets. 
The secrecy rate of each legitimate user is defined as the difference between the achievable rate at the intended receiver and the maximum rate achievable by any eavesdropper, ensuring secure communication at the physical layer. 
Hence, the secrecy rate for the $k$-th user is given by
\begin{equation}
    \label{eq.secrecy_rate}
    \!\!\!\text{SR}_k \!=\! \left[\log_2\!\left(1 \!+\! \rho^L_k\right) \!-\! \max_j \left(\log_2\!\left(1 \!+\! \rho^E_{k,j}\right)\right)\right]^+\!\!\!\!\!\!,\!\!
\end{equation}
where $\rho^L_k$ denotes the \ac{SINR} ratio at the $k$-th legitimate user, and $\rho^E_{k,j}$ represents the signal-to-noise ratio of the $k$-th user's signal as intercepted by the $j$-th eavesdropper. 
The operator $[\cdot]^+$ ensures that the secrecy rate remains non-negative. 

The \ac{SINR} at the $k$-th legitimate user is defined as \cite{SuTWC2021}
\begin{equation}
    \label{eq.SINR_B_k}
    \begin{aligned}
    \rho^L_k &= \frac{\mathbb{E} \left[ \left| \boldsymbol{h}_k \boldsymbol{w}_k s_k \right|^2 \right]}{
    \sum\limits_{i \ne k} \mathbb{E} \left[ \left| \boldsymbol{h}_k \boldsymbol{w}_i s_i \right|^2 \right] +
    \mathbb{E} \left[ \left| \boldsymbol{h}_k \boldsymbol{n}_{\text{eff}} \right|^2 \right] + \sigma_{z_k}^2} \\
    &= \frac{\boldsymbol{h}_k \boldsymbol{W}_k \boldsymbol{h}_k^{\text{H}} \|\boldsymbol{s}_k\|^2}{
    \sum\limits_{i \ne k} \boldsymbol{h}_k \boldsymbol{W}_i \boldsymbol{h}_k^{\text{H}} \|\boldsymbol{s}_i\|^2 +
    \boldsymbol{h}_k \boldsymbol{R}_{n_{\text{eff}}} \boldsymbol{h}_k^{\text{H}} + \sigma_{z_k}^2},
    \end{aligned}
\end{equation}
where $\boldsymbol{h}_k \in \mathbb{C}^{1 \times N_t}$ represents the channel vector from the transmitter to the $k$-th \ac{CU}, $\boldsymbol{W}_k = \boldsymbol{w}_k \boldsymbol{w}_k^{H}$ denotes the beamforming covariance matrix of a given user $k$, and $\boldsymbol{R}_{n_{\text{eff}}} \in \mathbb{C}^{N_t \times N_t}$ is the covariance matrix of the effective \ac{AN}. 
Moreover, the term \(\sigma_{z_k}^2\) represents the variance of the \ac{AWGN} at the $k$-th legitimate user's receiver.
Considering PSK-modulated symbols, we have $\|\boldsymbol{s}_k\|^2 = 1$.

The received \ac{SNR} at the $j$-th eavesdropping target, resulting from the $k$-th legitimate user's signal, is expressed as
\begin{equation}
    \label{eq.SINR_E_j}
    \rho^E_{k,j} = 
    \frac{|\alpha_j|^2 \boldsymbol{a}^{\text{H}}(\theta_j) \boldsymbol{W}_k \boldsymbol{a}(\theta_j)}{
    |\alpha_j|^2 \boldsymbol{a}^{\text{H}}(\theta_j) \boldsymbol{R}_{n_{\text{eff}}} \boldsymbol{a}(\theta_j) + \sigma_e^2},
\end{equation}
where $\alpha_j$ is the complex path gain from the transmitter to eavesdropper $E_j$, and $\boldsymbol{a}(\theta_j)$ is the array steering vector in the direction of eavesdropper $j$. 

Based on the above definitions, the total system secrecy rate can be defined as
\begin{equation}
    \label{eq.sum_secrecy_rate}
    \text{SR} \!\triangleq\! \sum_{k=1}^K \!\left[\log_2\!\left(1 \!+\! \rho^L_k\right) \!-\! \max_j \left(\log_2\!\left(1 \!+\! \rho^E_{k,j}\right)\right)\right]^{\!+}\!\!\!\!.
\end{equation}

This objective function seeks to optimize the overall confidentiality of the system by enhancing the \ac{SINR} at legitimate users while minimizing the potential \ac{SNR} leakage to any of the eavesdroppers. The worst-case eavesdropping scenario is considered via the $\max$ operator, thereby ensuring robustness in secure communication design.

\section{Optimization Formulation and \\ Fairness-Aware Strategy}
\label{sec:structuring_opt}

One of the key contributions of this work is the careful and strategic design of constraints for the proposed \ac{ISAC} system. 
Instead of relying on conventional formulations, we introduce a novel set of constraints that align with the system’s unique objectives and secrecy performance optimization. 
These constraints ensure that the intentional generation of \ac{AN} reinforces radar detection capabilities while avoiding interference with authorized \acp{CU} for maintaining communications quality. 
Additionally, power limitations and \ac{QoS} requirements are incorporated to guarantee system feasibility and to maximize the overall secrecy rate.
\subsection{Constraints on Communications Performance}

To prevent interference with the legitimate communication users, the \ac{AN} vector $\boldsymbol{n}_{\text{eff}}$ is designed to lie in the null space of the communication users' channel matrix. 
Let \( \boldsymbol{H} = [\boldsymbol{h}_1^{\text{T}}, \cdots, \boldsymbol{h}_k^{\text{T}}, \cdots, \boldsymbol{h}_K^{\text{T}}]^{\text{T}} \in \mathbb{C}^{K \times N_t} \) be the concatenated channel matrix of all communication users.
Then, the projected effective noise can be defined as
\begin{equation}
    \label{eq.n_eff}
    \boldsymbol{n}_{\text{eff}} = \boldsymbol{P}^\perp \boldsymbol{n},
\end{equation}
where $\boldsymbol{P}^\perp$ is the orthogonal projection matrix onto the null space of $\boldsymbol{H}$, which can be calculated as
\begin{equation}
\label{eq. p_c}
    \boldsymbol{P}^\perp = \boldsymbol{I}_{N_T} - \boldsymbol{H}^{\text{H}} (\boldsymbol{H} \boldsymbol{H}^{\text{H}})^{-1} \boldsymbol{H},
\end{equation}
with $ \boldsymbol{n}$ denoting the \ac{AN} vector to be optimized. Consequently, the covariance of the effective \ac{AN} becomes
\begin{equation}
    \label{eq.8}
    \boldsymbol{R}_{n_{\text{eff}}} = \boldsymbol{P}^\perp \boldsymbol{R}_n (\boldsymbol{P}^\perp)^{\text{H}},
\end{equation}
where $\boldsymbol{R}_n \triangleq \mathbb{E}[\boldsymbol{n} \boldsymbol{n}^{\text{H}}]$.
This guarantees that \ac{AN} does not interfere with the desired communication signals.

On the other hand, many existing works (e.g., \cite{SuTWC2021, DeligiannisTAES2018}) introduce the following constraint to ensure that the \ac{SINR} at each $k$-th legitimate user meets a minimum threshold $\gamma_k$, thereby guaranteeing the required communication \ac{QoS} for all users.
\begin{equation}
    \label{eq.8.2}
    \rho^L_k \geq \gamma_k, \quad \forall k.
\end{equation}

However, such predefined \ac{SINR} thresholds often lead to suboptimal and unfair resource allocation, as the beamforming vectors are typically optimized just to meet the minimum requirements, favoring users with stronger channels while marginalizing those with weaker links. 
This imbalance undermines fairness and reduces overall system utility in multi-user scenarios. 
To address this issue, we replace the fixed \ac{SINR} constraints with \ac{JFI} in \ref{Optimization_Problem}, a simple, interpretable, and normalized metric that effectively promotes equitable resource distribution, making it particularly suitable for multi-user wireless systems \cite{JainTECHREPORT1998}.

\subsection{Constraints on Sensing Performance}
To ensure the sensing functionality is preserved, the received power at each radar target must exceed a predefined threshold $\eta_j$. 
Let $\boldsymbol{a}(\theta_j)$ denote the steering vector corresponding to direction $\theta_j$, with the same structure as in equation \eqref{eq:steering_vector}. The received radar power at the $j$-th target is
\begin{equation}
    \label{eq.received_radar_power}
    P_j = \alpha_j \boldsymbol{a}^{\text{H}}(\theta_j) \underbrace{\left(  \sum\limits_{k=1}^K \boldsymbol{w}_k \boldsymbol{w}_k^{\text{H}} + \boldsymbol{R}_{n_{\text{eff}}} \right)}_{\tilde{\boldsymbol{W}}} \boldsymbol{a}(\theta_j).
\end{equation}

Thus, the radar power constraint at each target \(j\) becomes
\begin{equation}
    \label{eq.radar_power}
    \boldsymbol{a}^{\text{H}}(\theta_j) \tilde{\boldsymbol{W}} \boldsymbol{a}(\theta_j) \alpha_j \geq \eta_j.
\end{equation}

Motivated by the 3 dB main-beamwidth  design for MIMO radar in \cite{Li_2007}, we propose a method that constrains the main-beam width within a specified angular interval $\theta_0$, where $\theta_0$ is a pre-defined angle such that 2$\theta_0$ is equal to the desired 3-dB beam width of the waveform, given by
\begin{equation}
    \label{eq.radar_beam_width}
\boldsymbol{a}^{\text{H}}(\theta_j \pm \theta_0) \tilde{\boldsymbol{W}} \boldsymbol{a}(\theta_j \pm \theta_0) \alpha_j \leq \frac{\eta_j}{2}, \quad \forall j,
\end{equation}
where this constraint enforces a main beamwidth of $2\theta_0$, ensuring sufficient angular resolution for reliable radar target detection.

\subsection{Final Optimization Problem}
\label{Optimization_Problem}
Considering the design variables and constraints, the complete joint optimization problem can be formulated as
\begin{equation}
\label{eq.optimization_problem}
\begin{aligned}
&\max_{\boldsymbol{w}_k, \boldsymbol{n}} \; \text{SR} \!=\!\! \sum_{k=1}^K \!\!
\left[ \!
\log_2 \!\! \left(\!\!1 \!+\! 
\frac{\boldsymbol{h}_k \boldsymbol{w}_k \boldsymbol{w}_k^{\text{H}} \boldsymbol{h}_k^{\text{H}}}
{\sum\limits_{i \neq k} \! \boldsymbol{h}_k \boldsymbol{w}_i \boldsymbol{w}_i^{\text{H}} \boldsymbol{h}_k^{\text{H}} \!+\! 
\boldsymbol{h}_k \boldsymbol{n}_{\text{eff}} \boldsymbol{n}_{\text{eff}}^{\text{H}} \boldsymbol{h}_k^{\text{H}} \!+\! \sigma_k^2}
\!\!\right) \right. \\
& \left.
\hspace{8ex} - \max_j \log_2 \left(1 + 
\frac{\alpha_j^2 \boldsymbol{a}^{\text{H}}(\theta_j) \boldsymbol{w}_k \boldsymbol{w}_k^{\text{H}} \boldsymbol{a}(\theta_j)}
{\alpha_j^2 \boldsymbol{a}^{\text{H}}(\theta_j) \boldsymbol{n}_{\text{eff}} \boldsymbol{n}_{\text{eff}}^{\text{H}} \boldsymbol{a}(\theta_j) + \sigma_e^2}
\right)
\right]^+ \\
& \text{subject to} \quad \\
& \hspace{10ex} \text{(a)}\; \mathrm{Tr}(\boldsymbol{w}_k \boldsymbol{w}_k^{\text{H}}) \leq P_k, \quad \forall k, \\
& \hspace{10ex} \text{(b)}\; \mathrm{Tr}(\boldsymbol{R}_{n_{\text{eff}}}) + \sum_{k=1}^K P_k \leq P_A, \\
& \hspace{10ex} \text{(c)}\; \rho^L_k \geq \gamma_k, \quad \forall k, \\
& \hspace{10ex} \text{(d)}\; \boldsymbol{a}^{\text{H}}(\theta_j) \tilde{\boldsymbol{W}} \boldsymbol{a}(\theta_j) \alpha_j \geq \eta_j, \quad \forall j, \\
& \hspace{10ex} \text{(e)}\; \boldsymbol{a}^{\text{H}}(\theta_j \pm \theta_0) \tilde{\boldsymbol{W}} \boldsymbol{a}(\theta_j \pm \theta_0) \alpha_j \leq \frac{\eta_j}{2}, \quad \forall j. 
\end{aligned}
\end{equation}

This formulation ensures that the \ac{AN} is optimally projected to prevent interference with the communication functionality, while the sensing performance is preserved through power guarantees in desired directions, with the leakage to undesired directions controlled and the total transmit power remaining within system limits.

To address the non-smoothness and non-convexity introduced by the $[\,\cdot\,]^+$ operator and the $\max_j$ term in the original secrecy rate expression, we reformulate the problem by maximizing a weighted sum-rate of legitimate users subject to secrecy constraints that upper-bound the eavesdroppers' achievable rates.

This transformation converts the original intractable objective into a smooth and more tractable form, enabling the use of standard optimization techniques while ensuring a predefined level of secrecy.
The equivalent problem can then be cast as
\begin{equation}
\label{eq.13}
\begin{aligned}
&\hspace{-10ex} \max _{\boldsymbol{w}_k, \boldsymbol{n}_{\text{eff}}} \quad \sum_{k=1}^K \mu_k\log_2\!\!\left(\!\!1 \!+\! \frac {\boldsymbol{h}_k \boldsymbol{w}_k \boldsymbol{w}_k^{\text{H}} \boldsymbol{h}_k^{\text{H}}}
{\sum\limits_{i \neq k} \! \boldsymbol{h}_{k} \boldsymbol{w}_i \boldsymbol{w}_i^{\text{H}} \boldsymbol{h}_{k}^{\text{H}} \!+\! \boldsymbol{h}_k \boldsymbol{n}_{\text{eff}} \boldsymbol{n}_{\text{eff}}^{\text{H}} \boldsymbol{h}_k^{\text{H}} \!+\! \sigma_{k}^2}\!\! \right) \\
\text{subject to:} \\
& \hspace{-2ex}\text{(a)}\;  \log_2\!\left(\!\!1 + \frac {\alpha_j^2 \boldsymbol{a}^{\text{H}}(\theta_j) \boldsymbol{w}_k \boldsymbol{w}_k^{\text{H}} \boldsymbol{a}(\theta_j)}{\alpha_j^2 \boldsymbol{a}^{\text{H}}(\theta_j) \boldsymbol{n}_{\text{eff}} \boldsymbol{n}_{\text{eff}}^{\text{H}} \boldsymbol{a}(\theta_j) + \sigma_e^2}\!\!\right) \!\leq\! \beta_j, \;\forall j \\
& \hspace{-2ex}\text{(b)}\; \mathrm{Tr}(\boldsymbol{w}_k \boldsymbol{w}_k^{\text{H}}) \leq P_k,\quad \forall k, \\
& \hspace{-2ex}\text{(c)}\; \mathrm{Tr}(\boldsymbol{R}_{n_{\text{eff}}}) + \sum_{k=1}^K P_k \leq P_A, \\
& \hspace{-2ex}\text{(d)}\; \boldsymbol{a}^{\text{H}}(\theta_j) \tilde{\boldsymbol{W}} \boldsymbol{a}(\theta_j) \alpha_j \geq \eta_j, \quad \forall j, \\
& \hspace{-2ex}\text{(e)}\; \boldsymbol{a}^{\text{H}}(\theta_j \pm \theta_0) \tilde{\boldsymbol{W}} \boldsymbol{a}(\theta_j \pm \theta_0) \alpha_j \leq \frac{\eta_j}{2},\quad \forall j, 
\end{aligned}
\end{equation}
which can be iteratively divided into two optimization subproblems, one for optimizing over $\boldsymbol{w}_k$ and the other for optimizing over $\boldsymbol{n}$.

To circumvent the non-convexity of constraint (c) in \eqref{eq.optimization_problem}, we reformulate the original problem by introducing a fairness-aware metric based on Jain’s index, defined as
\begin{equation}
   F_{\mathrm{SINR}}(\boldsymbol{\mu}) \triangleq \frac{\left[\sum\limits_{k=1}^{K}\mu_k \cdot \rho^L_k\right]^2}{K \sum\limits_{k=1}^{K} \left( \mu_k \cdot \rho^L_k \right)^2}.
\end{equation}

This index satisfies \( \frac{1}{K} \leq F_{\mathrm{SINR}}(\boldsymbol{\mu}) \leq 1 \), where equality with 1 holds if and only if the \acp{SINR} are equal across users, i.e., perfectly fair allocation. 
To guarantee a minimum level of fairness, we impose the constraint \( F_{\mathrm{SINR}}(\boldsymbol{\mu}) \geq \xi_F \), where \( \xi_F \in (\frac{1}{K},1] \) is a design threshold.
We seek to determine the weight vector \( \boldsymbol{\mu} \in \mathbb{R}_+^K \) lying on the probability simplex \( \Delta = \{ \boldsymbol{\mu} \in \mathbb{R}_+^K: \sum_{k=1}^{K} \mu_k = 1 \} \), which balances throughput and fairness based on a trade-off parameter \( \chi \in [0, 1] \). 
The composite optimization objective is expressed as
\begin{equation}
\label{eq:fairness_tradeoff}
\max_{\boldsymbol{\mu} \in \Delta} \; (1 - \chi) \, G \sum_{k=1}^{K} \mu_k \log_2(1 + \rho^L_k) + \chi F_{\mathrm{SINR}}(\boldsymbol{\mu}),
\end{equation}
where \( G \) is a normalization constant defined to ensure that both terms in \eqref{eq:fairness_tradeoff} are normalized to the unit interval, i.e.,
\begin{equation}
    G \triangleq \left( \max_{\boldsymbol{\mu} \in \Delta} \sum_{k=1}^{K} \mu_k \log_2(1 + \rho^L_k) \right)^{-1}.
\end{equation}

\subsection{Closed-Form Solutions at Extreme Tradeoff Values}

We can identify two analytically tractable edge cases to obtain closed-form solutions.
\vspace{8pt}
\begin{itemize}
    \item \textit{Throughput Maximization (\( \chi = 0 \))}: The objective function reduces to a weighted sum-rate maximization, where the optimal solution places all the weight on the user with the highest \ac{SINR} (in the absence of regularization).
    \item \textit{Fairness Maximization (\( \chi = 1 \))}: The fairness metric \( F_{\mathrm{SINR}} \) is maximized when the effective \acp{SINR} are equalized, leading to the solution
    \vspace{-5pt}
    \begin{equation}
        \mu_k^\star = \frac{c}{\rho^L_k}, \quad \text{where} \quad c = \left( \sum_{k=1}^K \frac{1}{\rho^L_k} \right)^{-1}.
    \end{equation}
\end{itemize}
\vspace{-8pt}
\subsection{Entropy-Regularized Iterative Optimization Strategy}

To interpolate between the extreme regimes of fairness and throughput, we propose an iterative procedure over a discretized tradeoff path \( \{ \chi_t \}_{t=0}^T \), transitioning from full fairness (\( \chi_0 = 1 \)) to full throughput (\( \chi_T = 0 \)).
To ensure stability and explore smoother solutions, we incorporate entropy regularization into the objective as
\begin{equation}
\label{eq:composite_objective}
    \begin{aligned}
    \mathcal{L}(\boldsymbol{\mu}; \chi_t) &= (1 - \chi_t) G \sum_{k=1}^{K} \mu_k \log_2(1 + \rho_k^L)  \\
    & + \chi_t F_{\mathrm{SINR}}(\boldsymbol{\mu}) - \nu H(\boldsymbol{\mu}),
    \end{aligned}
\end{equation}
where \( H(\boldsymbol{\mu}) = -\sum_{k=1}^K \mu_k \log \mu_k \) denotes the Shannon entropy and \( \nu > 0 \) is a regularization parameter that encourages diversity in the weight allocation.
The parameter \( \nu \) is selected to balance smoothness and concentration. Typical values are empirically chosen to maintain solution sparsity while avoiding degeneracy.

The gradient of \( \mathcal{L} \) with respect to \( \mu_k \) is given by
\begin{equation}
\label{eq:gradient_mu}
\begin{aligned}
    \nabla_{\mu_k} \mathcal{L} &= (1 - \chi_t) G \log_2(1 + \rho_k^L) \\
    &\quad + \chi_t \left[ \frac{2(\boldsymbol{\mu}^T \boldsymbol{r}) r_k}{K \| \boldsymbol{\mu} \circ \boldsymbol{r} \|^2} - \frac{2(\boldsymbol{\mu}^T \boldsymbol{r})^2 r_k^2}{K \| \boldsymbol{\mu} \circ \boldsymbol{r} \|^4} \right] - \nu (1 + \log \mu_k),
\end{aligned}
\end{equation}
where \( \boldsymbol{r} = [\rho_1^L, \dots, \rho_K^L]^T \), and \( \circ \) denotes element-wise multiplication.
\vspace{-2ex}
\begin{algorithm}[H]
\caption{Enhanced H-FRO: Hybrid Fairness–Rate Optimization with Fairness Penalty and Entropy Regularization}
\label{alg:enhanced-HFRO}
\begin{algorithmic}[1]
\STATE \textbf{Input:} Initial weights \( \mu_k^{(0)} = \frac{1}{K} \), tradeoff path \( \{ \chi_t \}_{t=0}^T \), minimum fairness \( \xi_F \), entropy weight \( \nu > 0 \), penalty weight \( \lambda > 0 \), trust region rate \( \eta \in (0,1] \), inner iterations \( I \)
\STATE \textbf{Output:} Optimized weight vector \( \boldsymbol{\mu}^{(T)} \) and final sum secrecy rate
\vspace{5pt}
\STATE Calculate SINR vector \( \boldsymbol{\rho}^L \) via Algorithm~\ref{alg:NQT}.
\STATE Define simplex \( \Delta := \{ \boldsymbol{\mu} \in \mathbb{R}_+^K : \sum_k \mu_k = 1 \} \).
\STATE Initialize \( \mu_k^{(1)} \gets \frac{1 / \rho_k^L}{\sum_{k=1}^K 1 / \rho_k^L}, \quad \forall k \).

\FOR{$t = 1$ to $T$}
    \STATE Set \( \chi \gets \chi_t \), \( \boldsymbol{\mu} \gets \boldsymbol{\mu}^{(t-1)} \), \( \boldsymbol{\mu}_{\text{old}} \gets \boldsymbol{\mu} \).
    \FOR{$i = 1$ to $I$}
        \STATE Compute gradient \( \nabla_{\mu_k} \mathcal{L}_{\text{penalized}} \) via \eqref{eq:gradient_mu}.
        \STATE Update \( \boldsymbol{\mu} \gets \boldsymbol{\mu} + \alpha \cdot \nabla_{\boldsymbol{\mu}} \).
        \STATE Project \( \boldsymbol{\mu} \) onto simplex \( \Delta \).
        \STATE Apply trust region: \( \boldsymbol{\mu} \gets (1 - \eta)\boldsymbol{\mu}_{\text{old}} + \eta \boldsymbol{\mu} \).
    \ENDFOR
    \STATE Set \( \boldsymbol{\mu}^{(t)} \gets \boldsymbol{\mu} \).
\ENDFOR
\STATE \textbf{Return:} \( \boldsymbol{\mu}^{(T)} \)
\STATE Update secrecy rate using Algorithm~\ref{alg:NQT} with optimal \( \boldsymbol{\mu}^{(T)} \)
\end{algorithmic}
\end{algorithm}

To address the non-convex fairness constraint, we adopt a \textit{penalty-based relaxation} rather than hard projection. 
Specifically, the fairness constraint \( F_{\mathrm{SINR}} \geq \xi_F \) is encoded into the objective via a differentiable penalty defined by
\begin{equation}
    \mathcal{L}_\text{penalized}(\boldsymbol{\mu}; \chi_t) \triangleq \mathcal{L}(\boldsymbol{\mu}; \chi_t) - \lambda \cdot \left( \max(0, \xi_F - F_{\mathrm{SINR}}(\boldsymbol{\mu}) ) \right)^2,
\end{equation}
where \( \lambda > 0 \) is a penalty coefficient controlling constraint enforcement. 
This approach avoids projection onto a non-convex set and ensures differentiability.

The proposed approach integrates multiple convex components (entropy, throughput) with a non-convex fairness penalty, yielding a \textit{partially non-convex} landscape and is summarized in Algorithm \ref{alg:enhanced-HFRO}.
While global optimality is not guaranteed, the use of entropy regularization and gradual transition over \( \{ \chi_t \} \) promotes smooth convergence to locally optimal and practically fair solutions.

\section{Dual Optimization Framework}
\label{sec:dual_opt_frmework}
In this section, we develop a dual optimization framework to tackle the joint beamforming and noise design problem by decomposing it into two subproblems. The first subproblem optimizes the beamforming vectors assuming fixed noise, reformulating the original non-convex problem into a more tractable form via auxiliary variables and a dual \ac{QT}. This approach enables closed-form updates for key variables and leverages convex constraints to efficiently solve the problem. Subsequent subsections detail the problem formulation, the dual \ac{QT} application, and the derivation of optimal solutions for the introduced auxiliary variables.
\vspace{-10pt}
\subsection{Subproblem I: Non-homogeneous \ac{QT} for Beamforming}
Let us first consider the optimization over $\boldsymbol{w}_k$, under the assumption that $\boldsymbol{n}$ is fixed.
In addition, in order to make the optimization more tractable, we allocate the constraints (a) and (b) from \eqref{eq.13} to the first subproblem and the constraints (c-e) to the second subproblem.
Then, the equivalent first optimization problem can be expressed as
\vspace{-1ex}
\begin{equation}
\label{eq.14}
\begin{aligned}
&\max _{\boldsymbol{w}_k}  \sum_{k=1}^K \mu_k\log_2\left(1 + M_k(\boldsymbol{w}_k) \right) \\
&\text{subject to:} \\
& \hspace{3ex}\text{(a)}\;  \log_2\!\left(\!\!1 + \frac {\alpha_j^2 \boldsymbol{a}^{\text{H}}(\theta_j) \boldsymbol{w}_k \boldsymbol{w}_k^{\text{H}} \boldsymbol{a}(\theta_j)}{\alpha_j^2 \boldsymbol{a}^{\text{H}}(\theta_j) \boldsymbol{n}_{\text{eff}} \boldsymbol{n}_{\text{eff}}^{\text{H}} \boldsymbol{a}(\theta_j) + \sigma_e^2}\!\!\right) \!\leq\! \beta_j,\quad \forall j, \\
&\hspace{3ex}(b)\; \mathrm{Tr}(\boldsymbol{w}_k\boldsymbol{w}_k^{\text{H}}) \leq P_k, \quad \forall k, 
\end{aligned}
\end{equation}
where we use the definition
\begin{equation}
\label{eq.17}
   M_k(\boldsymbol{w}_k) \triangleq e_k^*(\boldsymbol{w}_k)B_k^{-1}(\boldsymbol{w}_k)e_k(\boldsymbol{w}_k),
\end{equation}
where $(\cdot)^*$ is the complex conjugate with \(e_k(\boldsymbol{w}_k) \triangleq \boldsymbol{h}_k \boldsymbol{w}_k \in \mathbb{C}\) and \(B_k(\boldsymbol{w}_k) \in \mathbb{C}\) denoting a covariance value that captures the interference from other users and noise effects given by
\begin{equation}
\label{eq.18}
   B_k(\boldsymbol{w}_k) = \sum_{\substack{i\neq k}} \boldsymbol{h}_{k} \boldsymbol{w}_i \boldsymbol{w}_i^{\text{H}} \boldsymbol{h}_{k}^{\text{H}} + \boldsymbol{h}_k \boldsymbol{n}_{\text{eff}} \boldsymbol{n}_{\text{eff}}^{\text{H}} \boldsymbol{h}_k^{\text{H}} + \sigma_{k}^2.
\end{equation}

Let us first note that the constraint (a) in equation \eqref{eq.14} can be expressed as a convex inequality if $\boldsymbol{n}$ is held fixed.
Therefore, the constraint
\begin{equation}
\label{eq.15}
\log_2\left(1 + \frac {\alpha_j^2 \boldsymbol{a}^{\text{H}}(\theta_j) \boldsymbol{w}_k \boldsymbol{w}_k^{\text{H}} \boldsymbol{a}(\theta_j)}{\alpha_j^2 \boldsymbol{a}^{\text{H}}(\theta_j) \boldsymbol{n}_{\text{eff}} \boldsymbol{n}_{\text{eff}}^{\text{H}} \boldsymbol{a}(\theta_j) + \sigma_e^2}\right) \leq \beta_j,
\end{equation}
can be equivalently expressed as
\begin{equation}
\label{eq.16}
\boldsymbol{w}_k^{\text{H}} \boldsymbol{a}(\theta_j) \boldsymbol{a}^{\text{H}}(\theta_j) \boldsymbol{w}_k \!\leq\! \frac{(2^{\beta_j} \!-\! 1) \!\! \left( \alpha_j^2 \boldsymbol{a}^{\text{H}}(\theta_j) \boldsymbol{n}_{\text{eff}} \boldsymbol{n}_{\text{eff}}^{\text{H}} \boldsymbol{a}(\theta_j) \!+\! \sigma_e^2 \right)}{\alpha_j^2}.
\end{equation}

Leveraging the above, the main sub-optimization portrayed in equation \eqref{eq.14} can be reformulated as follows \cite{ShenTSP2018, II-ShenTSP2018}.
\begin{align}
h(\boldsymbol{w}_k,\boldsymbol{\zeta}) &= \sum^K_{k=1} \mu_k \bigg[ \log_2 \,(1+\zeta_k) - \zeta_k \nonumber \\
&\hspace{-9ex} + (1\!+\!\zeta_k) e^*_k (\boldsymbol{w}_k) \left( e_k(\boldsymbol{w}_k) e^*_k(\boldsymbol{w}_k) \!+\! B_k(\boldsymbol{w}_k) \right)^{-1}\!\! e_k(\boldsymbol{w}_k) \bigg], \label{eq.19}
\end{align}
where $\bm{\zeta} \triangleq \{\zeta_1, \cdots, \zeta_K\}$ denotes a set of auxiliary variables.

Let us now also define \(\hat{B}_k(\boldsymbol{w}_k) \triangleq e_k(\boldsymbol{w}_k) e^* _k(\boldsymbol{w}_k) + B_k(\boldsymbol{w}_k)\) and \(\hat{M}_k(\boldsymbol{w}_k) \triangleq e^*_k (\boldsymbol{w}_k) \hat{B}_k^{-1}(\boldsymbol{w}_k) e_k(\boldsymbol{w}_k) \).

Leveraging the above definitions, the optimization problem defined in equation \eqref{eq.14} can be reformulated as
\begin{equation}
\label{eq.21}
\begin{aligned}
&\max _{\boldsymbol{w}_k,\bm{\zeta}} \;  h(\boldsymbol{w}_k,\boldsymbol{\zeta}) \!=\! \sum_{k=1}^{K} \mu_k \! \left[(1\!+\!\zeta_k)\hat{M}_k(\boldsymbol{w}_k) \!+\! \log_2(1\!+\!\zeta_k) \!-\! \zeta_k \right] \\
&\text{subject to:} \\
&\hspace{4ex} \text{(a)}\;  \log_2\!\left(\!\!1 + \frac {\alpha_j^2 \boldsymbol{a}^{\text{H}}(\theta_j) \boldsymbol{w}_k \boldsymbol{w}_k^{\text{H}} \boldsymbol{a}(\theta_j)}{\alpha_j^2 \boldsymbol{a}^{\text{H}}(\theta_j) \boldsymbol{n}_{\text{eff}} \boldsymbol{n}_{\text{eff}}^{\text{H}} \boldsymbol{a}(\theta_j) + \sigma_e^2}\!\!\right) \!\leq\! \beta_j,\quad \forall j, \\
&\hspace{4ex}(b)\; \mathrm{Tr}(\boldsymbol{w}_k\boldsymbol{w}_k^{\text{H}}) \leq P_k, \quad\forall k.
\end{aligned}
\end{equation}

Next, by utilizing a dual complex \ac{QT}, we can re-express $\hat{M}_k(\boldsymbol{w}_k)$ as 
\begin{equation}
\label{eq.20}
\hat{M}_k(\boldsymbol{w}_k) = 2{\operatorname{Re}}\left\{ y_k^* e_k (\boldsymbol{w}_k)\right\} -y_k^* \hat{B}_k(\boldsymbol{w}_k)y_k,
\end{equation}
where $y_k \in \mathbb{C}$ is an auxiliary variable.

Substituting the definition of $\hat{B}_k(\boldsymbol{w}_k)$ into the second term of equation \eqref{eq.20}, one can further evaluate that
\begin{equation}
\label{eq.23}
\begin{aligned}
y_k^* \hat{B}_k(\boldsymbol{w}_k)y_k &= \boldsymbol w_k^{\text{H}} \left(\boldsymbol h_k^{\text{H}} y_k y_k^* \boldsymbol h_k + \sum_{\substack{i\neq k}} \boldsymbol h_k^{\text{H}} y_i y_i^* \boldsymbol h_{k}\right) \boldsymbol w_k \\
&+ \boldsymbol{n}^{\text{H}} (\boldsymbol{P}^\perp)^{\text{H}} \boldsymbol h_k^{\text{H}} y_k y_k^* \boldsymbol h_k \boldsymbol{P}^\perp \boldsymbol{n} + \sigma_k^2 y_k^* y_k.
\end{aligned}
\end{equation}

To address the optimization efficiently, we adopt the non-homogeneous quadratic transform, a recent advancement in \ac{FP} that outperforms the classical \ac{QT} in convergence and efficiency \cite{ShenJSAC2024}. Inspired by the non-homogeneous bounding approach from majorization-minimization theory \cite{Sun_TCP2017}, this method eliminates costly matrix inverse operations, making it particularly suitable for large-scale wireless systems with massive antenna arrays. Therefore, to derive an upper bound for this expression, we apply the upper-bounding identity as follows
\begin{equation}
\label{eq.24}
\begin{aligned}
    &\boldsymbol w_k^{\text{H}} \bigg(\boldsymbol h_k^{\text{H}} y_k y_k^* \boldsymbol h_k + \sum_{\substack{i\neq k}} \boldsymbol h_k^{\text{H}} y_i y_i^* \boldsymbol h_{k}\bigg) \boldsymbol w_k \\
    &\leq \kappa_k \boldsymbol{w}_k^{\text{H}} \boldsymbol{w}_k + 2{\operatorname{Re}}\bigg\{\boldsymbol{w}_k^{\text{H}}\bigg( \Big(\boldsymbol h_k^{\text{H}} y_k y_k^* \boldsymbol h_k + \sum_{\substack{i\neq k}} \boldsymbol h_k^{\text{H}} y_i  y_i^* \boldsymbol h_{k}\Big) \\
    &-\kappa_k\boldsymbol I\bigg)\boldsymbol{z}_k\bigg\} + \boldsymbol{z}_k^{\text{H}} \bigg(\!\kappa_k\boldsymbol I \!-\! \Big(\boldsymbol h_k^{\text{H}} y_k y_k^* \boldsymbol h_k \!+\! \sum_{\substack{i\neq k}} \boldsymbol h_k^{\text{H}} y_i y_i^* \boldsymbol h_{k}\Big)\!\bigg)\boldsymbol{z}_k,
\end{aligned}
\end{equation}
with $\boldsymbol{z}_k \in \mathbb{C}^{N_t \times 1}$ being an auxiliary variable and $\kappa_k$ is a tunable hyperparameter that can be chosen such that 
\[
\kappa_k \geq \kappa_{\text{max}}(\boldsymbol{D}_k),
\]
in which $\kappa_{\text{max}}(\boldsymbol{D}_k)$ denotes the maximum eigenvalue of the matrix $\boldsymbol{D}_k \in \mathbb{C}^{N_t \times N_t}$ defined as
\begin{equation}
\label{eq.27}
\boldsymbol{D}_k = \mu_k(1+\zeta_k)\boldsymbol{h}_k^{\text{H}} y_k y_k^* \boldsymbol{h}_k + \sum_{\substack{i\neq k}} \mu_i(1+\zeta_i)\boldsymbol{h}_k^{\text{H}} y_i y_i^* \boldsymbol{h}_{k}.
\end{equation}
It should be noted that equality holds if 
\begin{equation}
 \boldsymbol w_k = \boldsymbol z_k.
\end{equation}

This leads us to the equivalent quadratic dual for the optimization problem in equation \eqref{eq.21}, given by
\begin{equation}
\label{eq.25}
    \begin{aligned}
    f_k(\boldsymbol{w}_k, y_k, \boldsymbol{z}_k, \zeta_k) 
    &\!=\!\! \sum_{k=1}^{K} \!\mu_k(1\!+\!\zeta_k)  \!\Bigg(\!\! 2 \operatorname{Re} \big\{\! y_k^* \boldsymbol{h_k}\boldsymbol w_k \!\big\} \!-\! \bigg(\!\! \kappa_k \boldsymbol{w}_k^{\text{H}} \boldsymbol{w}_k \\
    &\hspace{-16.5ex}+ 2{\operatorname{Re}}\bigg\{\boldsymbol{w}_k^{\text{H}}\bigg( \Big(\boldsymbol h_k^{\text{H}} y_k y_k^* \boldsymbol h_k + \sum_{\substack{i\neq k}} \boldsymbol h_k^{\text{H}} y_i y_i^* \boldsymbol h_{k}\Big)-\kappa_k\boldsymbol I_{N_t}\bigg)\boldsymbol{z}_k\bigg\} \\
   &\hspace{-16.5ex}+ \boldsymbol{z}_k^{\text{H}} \bigg(\kappa_k\boldsymbol I_{N_t} - \Big(\boldsymbol h_k^{\text{H}} y_k y_k^* \boldsymbol h_k + \sum_{\substack{i\neq k}} \boldsymbol h_k^{\text{H}} y_i y_i^* \boldsymbol h_{k}\Big)\bigg)\boldsymbol{z}_k \\
   &\hspace{-16.5ex}+ \boldsymbol{n}^{\text{H}} (\boldsymbol{P}^\perp)^{\text{H}} \boldsymbol{h}_k^{\text{H}} y_k y_k^* \boldsymbol{h}_k \boldsymbol{P}^\perp \boldsymbol{n} \\
   &\hspace{-16.5ex}+ \sigma_k^2 y_k^* y_k \bigg) \Bigg)  + \sum_{k=1}^{K} \mu_k \left[ \log_2(1+\zeta_k) - \zeta_k \right].
\end{aligned}
\end{equation}

Simplifying equation \eqref{eq.25} using equation \eqref{eq.27} results in
\begin{equation}
\label{eq.26}
\begin{aligned}
    f_k(\boldsymbol{w}_k, y_k, \boldsymbol{z}_k, \zeta_k) \\
    &\hspace{-16ex}= \sum_{k=1}^{K}  \Bigg( 2 \operatorname{Re} \left\{ \mu_k(1+\zeta_k)y_k^* \boldsymbol{h}_k \boldsymbol{w}_k + \boldsymbol{w}_k^{\text{H}}\left( \kappa_k \boldsymbol{I}_{N_t} - \boldsymbol{D}_k\right)\boldsymbol{z}_k \right\} \\
    &\hspace{-16ex}+ \boldsymbol{z}_k^{\text{H}} \left( \boldsymbol{D}_k - \kappa_k \boldsymbol{I}_{N_t}\right)\boldsymbol{z}_k  - \kappa_k \boldsymbol{w}_k^{\text{H}} \boldsymbol{w}_k \\
    &\hspace{-16ex}- \mu_k(1+\zeta_k)\boldsymbol{n}^{\text{H}} (\boldsymbol{P}^\perp)^{\text{H}} \boldsymbol{h}_k^{\text{H}} y_k y_k^* \boldsymbol{h}_k \boldsymbol{P}^\perp \boldsymbol{n} \\
    &\hspace{-16ex}- \mu_k(1+\zeta_k)\sigma_k^2 y_k^* y_k 
   \Bigg)  + \sum_{k=1}^{K} \mu_k \left[ \log_2(1+\zeta_k) - \zeta_k \right].
\end{aligned}
\end{equation}
To solve this optimization problem, we initialize by setting 
\begin{equation}
    \label{eq.27.1}
    \boldsymbol{z}_k^{\star} = \boldsymbol{w}_k.
\end{equation}

Next, we derive the closed-form optimum of \(y_k\), for all \(k\). 
For this process, we treat \(\boldsymbol{w}_k\) and \(\bm{\zeta}\) as constants while optimizing with respect to \(y_k\). 
This allows us to focus on the terms in \(f_k(\boldsymbol{w}_k, y_k, \boldsymbol{z}_k, \zeta_k)\) that involve \(\boldsymbol{y}\), given by
\begin{equation}
\label{eq.28}
\begin{aligned}
    f_k(\boldsymbol{y})=& \sum_{k=1}^{K} \mu_k (1 + \zeta_k) \\
    & \times\left( 2 \operatorname{Re} \left\{ y_k^* e_k(\boldsymbol{w}_k) \right\} - y_k^* \hat{B}_k(\boldsymbol{w}_k) y_k \right)+c_k,
\end{aligned}
\end{equation}
where \(c_k\) is a constant independent of \(\boldsymbol{y}\).
Leveraging the latter, the optimal auxiliary variable \(y_k^{\star}\) is given by

\begin{equation}
\label{eq.29}
y_k^{\star} = \hat{B}_k^{-1}(\boldsymbol{w}_k) e_k(\boldsymbol{w}_k).
\end{equation}

After substituting \(\boldsymbol{z}_{k}^{\star}\) and \(y_{k}^{\star}\) into \(f_k(\boldsymbol{w}_k, y_k, \boldsymbol{z}_k, \zeta_k)\), we can solve for the optimal \(\zeta_{k}\) by setting
\[
\frac{\partial f_k(\boldsymbol{w}_k, y_k, \boldsymbol{z}_k, \zeta_k)}{\partial \zeta_{k}} = 0.
\]

This, in turn, yields
\begin{equation}
\label{eq.30}
\zeta_k^* = \frac{1}{\ln(2) \left( 1 + \tilde{n}_k + \sigma_k^2 y_k y_k^* - \operatorname{Re}  (y_k e_k(\boldsymbol{w}_k))  \right)} - 1.
\end{equation}
where $\tilde{n}_k \triangleq \boldsymbol{n}^{\text{H}} (\boldsymbol{P}^\perp)^{\text{H}} \boldsymbol{h}_k^{\text{H}} y_k^* y_k \boldsymbol{h}_k \boldsymbol{P}^\perp \boldsymbol{n}$.

Finally, we can express the updated weight \({\boldsymbol{w}}_{k}^{\star}\) as
\begin{equation}
\label{eq.31}
{\boldsymbol{w}}_{k}^{\star} = \boldsymbol{z}_{k}^{\star} + \frac{1}{\kappa_{k}}\left(\mu_{k}\left(1+\zeta_{k}^{\star}\right) \boldsymbol{h}_{k}^{\mathrm{H}} y_{k}^{\star} -\boldsymbol{D}_{k} \boldsymbol{z}_{k}^{\star} \right),
\end{equation}
with the $n$-th iteration given by
\begin{equation}
\label{eq.32}
\hat{\boldsymbol{w}}_{k}^{(n)} = \mathcal{P}_{\boldsymbol{w}_{k} \in \mathcal{W}_{k}} \left(\boldsymbol{z}_{k}^{\star} + \frac{1}{\kappa_{k}}\left(\mu_{k}\left(1+\zeta_{k}^{\star}\right) \boldsymbol{h}_{k}^{\mathrm{H}} y_{k}^{\star} -\boldsymbol{D}_{k} \boldsymbol{z}_{k}^{\star} \right)\right),
\end{equation}
where \(\mathcal{P}_{\boldsymbol{w}_{k} \in \mathcal{W}_{k}}\) denotes projection onto the feasible set \(\mathcal{W}_{k}\).

\subsection{Subproblem II: Non-homogeneous \ac{QT} for \ac{AN} Design}

Next, let us consider the optimization over $\boldsymbol{n}$, under the assumption that $\boldsymbol{w}_k$ is fixed.
Then, the resulting objective function to maximize is given by
\begin{equation}
\label{eq.33}
\begin{aligned}
&\hspace{-10ex}\max _{\boldsymbol{n}} \quad \sum_{k=1}^K \mu_k\log_2\!\!\left(\!\!1 \!+\! \frac {\boldsymbol{h}_k \boldsymbol{w}_k \boldsymbol{w}_k^{\text{H}} \boldsymbol{h}_k^{\text{H}}}
{\sum\limits_{i \neq k} \! \boldsymbol{h}_{k} \boldsymbol{w}_i \boldsymbol{w}_i^{\text{H}} \boldsymbol{h}_{k}^{\text{H}} \!+\! \boldsymbol{h}_k \boldsymbol{n}_{\text{eff}} \boldsymbol{n}_{\text{eff}}^{\text{H}} \boldsymbol{h}_k^{\text{H}} \!+\! \sigma_{k}^2}\!\! \right)\\
\text{subject to:} \\
&(a)\;\mathrm{Tr}(\boldsymbol{P}^\perp \boldsymbol{n} \boldsymbol{n}^{\text{H}}(\boldsymbol{P}^\perp)^{\text{H}}) + \sum_{k=1}^K P_k\leq P_{A}, \\
& \text{(b)}\; \boldsymbol{a}^{\text{H}}(\theta_j) \tilde{\boldsymbol{W}} \boldsymbol{a}(\theta_j) \alpha_j \geq \eta_j, \quad\forall j, \\
& \text{(c)}\; \boldsymbol{a}^{\text{H}}(\theta_j \pm \theta_0) \tilde{\boldsymbol{W}} \boldsymbol{a}(\theta_j \pm \theta_0) \alpha_j \leq \frac{\eta_j}{2},\quad\forall j.
\end{aligned}
\end{equation}

Leveraging a procedure similar to the previous sub-optimization, the objective function with respect to $\boldsymbol{n}$ can be equivalently stated as
\begin{align} 
\label{eq.36}
    h(\boldsymbol{n},\tilde{\bm{\zeta}}) &= \sum^K_{k=1}\tilde{\mu}_k \bigg[ \log_2 \,(1+\tilde{\zeta}_k ) -  \tilde{\zeta}_k \nonumber \\
    &+ (1+\tilde{\zeta}_k)e_k^*  (e_k\boldsymbol e_k^* + {B}_k(\boldsymbol{n}))^{-1}e_k \bigg],
\end{align}
where $\tilde{\bm{\zeta}} \triangleq \{\tilde{\zeta}_1, \cdots, \tilde{\zeta}_K\}$ denotes a set of auxiliary variables and we drop the inherent dependance of $\boldsymbol{w}_k$ on $e_k(\boldsymbol{w}_k)$ and hereafter denote it as $e_k$.
In addition, ${B}_k(\boldsymbol{n})$ is identical to ${B}_k(\boldsymbol{w}_k)$ defined in equation \eqref{eq.18} with the implicit dependence on $\boldsymbol{n}$ highlighted instead.

Next, by letting $M_k(\boldsymbol{n})$ to be identical to $M_k(\boldsymbol{w}_k)$ in equation \eqref{eq.17} with the emphasis on $\boldsymbol{n}$, we can define $\hat{B}_k(\boldsymbol{n}) \triangleq e_k e^*_k + B_k(\boldsymbol{n})$ and $\hat{M}_k(\boldsymbol{n})\triangleq e^*_k (e_k e^* _k + B_k(\boldsymbol{n}))^{-1}e_k$.
Based on the dual complex \ac{QT}, we can also rewrite $e_k^* \hat{B}_k^{-1}(\boldsymbol{n})e_k $ as $2{\operatorname{Re}}\left\lbrace \tilde{y}_k^* e_k \right\rbrace -\tilde{y}_k^* \hat{B}_k(\boldsymbol{n})\tilde{y}_k$, where $\tilde{y}_k \in \mathbb{C}$ is an auxiliary variable, which yields the optimization problem
\vspace{-10pt}
\begin{equation}
\label{eq.37}
\begin{aligned}
&\hspace{-10.5ex}\max_{\boldsymbol{n}, \tilde{\bm{\zeta}}}\; h(\boldsymbol{n},\tilde{\bm{\zeta}}) = \sum^K_{k=1}\tilde{\mu}_k \bigg[ \log_2 \,(1+\tilde{\zeta}_k ) -  \tilde{\zeta}_k + (1+\tilde{\zeta}_k)\hat{M}_k(\boldsymbol{n}) \bigg]\\
\text{subject to:} \\
&(a)\;\mathrm{Tr}(\boldsymbol{P}^\perp \boldsymbol{n} \boldsymbol{n}^{\text{H}}(\boldsymbol{P}^\perp)^{\text{H}}) + \sum_{k=1}^K P_k\leq P_{A}, \\
& \text{(b)}\; \boldsymbol{a}^{\text{H}}(\theta_j) \tilde{\boldsymbol{W}} \boldsymbol{a}(\theta_j) \alpha_j \geq \eta_j, \quad\forall j, \\
& \text{(c)}\; \boldsymbol{a}^{\text{H}}(\theta_j \pm \theta_0) \tilde{\boldsymbol{W}} \boldsymbol{a}(\theta_j \pm \theta_0) \alpha_j \leq \frac{\eta_j}{2}, \quad \forall j,
\end{aligned}
\end{equation}
\vspace{-15pt}
where the objective function can be rewritten as 
\begin{align}
\label{eq.38}
    f_r(\boldsymbol{n},\tilde{y}_k, \tilde{\bm{\zeta}}) &= \sum_{k=1}^{K} \tilde{\mu}_k\bigg[(1+\tilde{\zeta}_k)\left(2{\operatorname{Re}}\left\lbrace \tilde{y}_k^* e_k \right\rbrace -\tilde{y}_k^* \hat{B}_k(\boldsymbol{n})\tilde{y}_k \right) \nonumber \\
    &\hspace{11ex} +  \log_2 (1+\tilde{\zeta}_k) - \tilde{\zeta}_k \bigg].
\end{align}
Next, following equation \eqref{eq.23}, we have
\vspace{-10pt}
\begin{equation}
\label{eq.39}
\begin{aligned}
y_k^* \hat{B}_k(\boldsymbol{n})y_k &= \boldsymbol w_k^{\text{H}} \left(\boldsymbol h_k^{\text{H}} y_k y_k^* \boldsymbol h_k + \sum_{\substack{i\neq k}} \boldsymbol h_k^{\text{H}} y_i y_i^* \boldsymbol h_{k}\right) \boldsymbol w_k \\
&+ \boldsymbol{n}^{\text{H}} (\boldsymbol{P}^\perp)^{\text{H}} \boldsymbol h_k^{\text{H}} y_k y_k^* \boldsymbol h_k \boldsymbol{P}^\perp \boldsymbol{n} + \sigma_k^2 y_k^* y_k.
\end{aligned}
\end{equation}

In addition, inspired by the non-homogeneous bounding approach \cite{Sun_TCP2017}, the second term in equation \eqref{eq.39} can be upper bounded as 
\begin{equation}
\label{eq.41}
\begin{aligned}
    &\boldsymbol{n}^{\text{H}} \left((\boldsymbol{P}^\perp)^{\text{H}} \boldsymbol h_k^{\text{H}} \tilde{y}_k \tilde{y}_k^* \boldsymbol h_k \boldsymbol{P}^\perp \right)\boldsymbol{n}\\
   &\leq \tilde{\kappa}_k \boldsymbol{n}^{\text{H}} \boldsymbol{n} + 2{\operatorname{Re}}\left\{\boldsymbol{n}^{\text{H}}\left( \left((\boldsymbol{P}^\perp)^{\text{H}} \boldsymbol h_k^{\text{H}} \tilde{y}_k \tilde{y}_k^* \boldsymbol h_k \boldsymbol{P}^\perp \right)-\tilde{\kappa}_k\boldsymbol I\right)\boldsymbol{z}\right\} \\
   &+ \boldsymbol{z}^{\text{H}} \left(\tilde{\kappa}_k\boldsymbol I- \left((\boldsymbol{P}^\perp)^{\text{H}} \boldsymbol h_k^{\text{H}} \tilde{y}_k \tilde{y}_k^* \boldsymbol h_k \boldsymbol{P}^\perp \right)\right)\boldsymbol{z},
\end{aligned}
\end{equation}
where $\boldsymbol{z} \in \mathbb{C}^{N_t \times 1}$ is an auxiliary variable, and the equality holds if \(\boldsymbol{n} = \boldsymbol{z}\). $\tilde{\kappa}_k$ is a tunable hyperparameter that can be chosen such that 
\vspace{-10pt}
\[
\tilde{\kappa}_k \geq \tilde{\kappa}_{\text{max}}(\tilde{\boldsymbol{D}}_k),
\]
in which $ \tilde{\kappa}_{\text{max}}(\tilde{\boldsymbol{D}}_k)$ denotes the maximum eigenvalue of the matrix $\tilde{\boldsymbol{D}}_k \in \mathbb{C}^{N_t \times N_t}$ defined as
\begin{equation}
\label{eq.dtilde}
\tilde{\boldsymbol{D}}_k \triangleq \tilde{\mu}_k(1+\tilde{\zeta}_k)\left((\boldsymbol{P}^\perp)^{\text{H}} \boldsymbol{h}_k^{\text{H}} \tilde{y}_k \tilde{y}_k^* \boldsymbol{h}_k
\boldsymbol{P}^\perp \right).
\end{equation}
\vspace{-10pt}
Therefore, the equivalent quadratic dual for the optimization problem defined in equation \eqref{eq.37} becomes
\begin{equation}
\label{eq.42}
\begin{aligned}
    &f_r(\boldsymbol{n}, \tilde{y}_k, \boldsymbol{z}, \tilde{\bm{\zeta}}) = \sum_{k=1}^{K} \tilde{\mu}_k(1+\tilde{\zeta}_k)  \Bigg( 2 \operatorname{Re} \left\{ \tilde{y}_k^* \boldsymbol{h_k}\boldsymbol w_k \right\} \!-\! \bigg( \tilde{\kappa}_k \boldsymbol{n}^{\text{H}} \boldsymbol{n}\\
     & + 2{\operatorname{Re}}\left\{\boldsymbol{n}^{\text{H}}\left( \left((\boldsymbol{P}^\perp)^{\text{H}} \boldsymbol h_k^{\text{H}} \tilde{y}_k \tilde{y}_k^* \boldsymbol h_k \boldsymbol{P}^\perp \right)-\tilde{\kappa}_k\boldsymbol I\right)\boldsymbol{z}\right\} \\
   &+ \boldsymbol{z}^{\text{H}} \left(\tilde{\kappa}_k\boldsymbol I- \left((\boldsymbol{P}^\perp)^{\text{H}} \boldsymbol h_k^{\text{H}} \tilde{y}_k \tilde{y}_k^* \boldsymbol h_k \boldsymbol{P}^\perp \right)\right)\boldsymbol{z}\\
   &+ \boldsymbol w_k^{\text{H}} \bigg(\boldsymbol h_k^{\text{H}} \tilde{y}_k \tilde{y}_k^* \boldsymbol h_k + \sum_{\substack{i\neq k}} \boldsymbol h_k^{\text{H}} \tilde{y}_i \tilde{y}_i^* \boldsymbol h_{k}\bigg) \boldsymbol w_k  + \sigma_k^2 \tilde{y}_k^* \tilde{y}_k\bigg) \Bigg) \\
   &+ \sum_{k=1}^{K} \tilde{\mu}_k \left[  \log_2(1+\tilde{\zeta}_k) - \tilde{\zeta}_k \right].
\end{aligned}
\end{equation}

Next, simplifying equation \eqref{eq.42} using equation \eqref{eq.dtilde} yields
\begin{equation}
\label{eq.43}
\begin{aligned}
    &f_r(\boldsymbol{n}, \tilde{y}_k, \boldsymbol{z}, \tilde{\bm{\zeta}}) \\ 
    &= \sum_{k=1}^{K}  \Bigg( 2 \operatorname{Re} \left\{ \tilde{\mu}_k(1+\tilde{\zeta}_k)\tilde{y}_k^* \boldsymbol{h_k}\boldsymbol w_k + \boldsymbol{n}^{\text{H}}\left( \tilde{\kappa}_k\boldsymbol I - \tilde{\boldsymbol{D}}_k\right)\boldsymbol{z} \right\} \\
   &+ \boldsymbol{z}^{\text{H}} \left( \tilde{\boldsymbol{D}}_k -\tilde{\kappa}_k\boldsymbol I\right)\boldsymbol{z}  -\tilde{\kappa}_k \boldsymbol{n}^{\text{H}} \boldsymbol{n}  \\
   &- \tilde{\mu}_k(1+\tilde{\zeta}_k)\boldsymbol w_k^{\text{H}} \bigg(\boldsymbol h_k^{\text{H}} \tilde{y}_k \tilde{y}_k^* \boldsymbol h_k + \sum_{\substack{i\neq k}} \boldsymbol h_k^{\text{H}} \tilde{y}_i \tilde{y}_i^{\text{H}} \boldsymbol h_{k}\bigg) \boldsymbol w_k \\
   &-\tilde{\mu}_k(1+\tilde{\zeta}_k)\sigma_k^2 \tilde{y}_k^* \tilde{y}_k 
   \Bigg)  + \sum_{k=1}^{K} \tilde{\mu}_k \left[  \log_2(1+\tilde{\zeta}_k) - \tilde{\zeta}_k \right].
\end{aligned}
\end{equation}

To solve this optimization problem, we first set
\begin{equation}
\label{eq.43.1}
\boldsymbol z^{\star} = \boldsymbol{n}.
\end{equation}

Next, to compute the closed-form optimal value of \(\tilde{y}_k\), we fix \(\boldsymbol{n}\) and \(\tilde{\zeta}_k\) to optimize with respect to \(\tilde{y}_k\), treating the other variables as constants. 
Given this setup, we can proceed by focusing solely on the terms in \(f_r(\boldsymbol{n}, \tilde{y}_k, \boldsymbol{z}, \tilde{\bm{\zeta}})\) that involve \(\tilde{y}_k\) to find the closed-form expression for the optimal \(\tilde{y}_k^\star\).

The objective function with respect to \(\tilde{y}_k\) can be expressed as
\begin{equation}
\label{eq.45}
f_r(\tilde{y}_k) = \tilde{\mu}_k (1 + \tilde{\zeta}_k) \left( 2 \operatorname{Re} \left\{ \tilde{y}_k^* e_k  \right\} - \tilde{y}_k^* \hat{B}_k(\boldsymbol{n}) \tilde{y}_k \right)+c_r,
\end{equation}
where \(c_r\) is a constant independent of \(\tilde{y}_k\).

Therefore, the optimal auxiliary variable \(\tilde{y}_k^\star\) is given by
\begin{equation}
\label{eq.46}
   \tilde{y}_k^\star = \hat{B}_k^{-1}(\boldsymbol{n}) e_k.
\end{equation}

Substituting $\boldsymbol{z}^{\star}$ and $\tilde{y}_k^{\star}$ into $f_r(\boldsymbol{n}, \tilde{y}_k, \boldsymbol{z}, \tilde{\zeta}_k)$, we can solve for the optimal $\tilde{\zeta}_k^\star$ by solving
\[
\frac{\partial f_r(\boldsymbol{n}, \tilde{y}_k, \boldsymbol{z}, \tilde{\zeta}_k)}{\partial \tilde{\zeta}_k} = 0.
\]

Taking the derivative with respect to \(\tilde{\zeta}_k\), we obtain
\begin{equation}
\label{eq.47}
\begin{aligned}
    \frac{\partial f_r(\tilde{\zeta}_k)}{\partial \tilde{\zeta}_k} 
    &= \tilde{\mu}_k \left( 2 \operatorname{Re} \left\{ \tilde{y}_k^* \boldsymbol{h}_k \boldsymbol{w}_k \right\} \right) \\
    &\quad - \tilde{\mu}_k \boldsymbol{w}_k^{\text{H}} \left( \boldsymbol{h}_k^{\text{H}} \tilde{y}_k \tilde{y}_k^* \boldsymbol{h}_k + \sum_{i \neq k} \boldsymbol{h}_k^{\text{H}} \tilde{y}_i \tilde{y}_i^{\text{H}} \boldsymbol{h}_k \right) \boldsymbol{w}_k \\
    &\quad - \tilde{\mu}_k \sigma_k^2 \tilde{y}_k^* \tilde{y}_k 
    + \frac{\tilde{\mu}_k}{\ln(2)} \cdot \frac{1}{1 + \tilde{\zeta}_k} - \tilde{\mu}_k = 0.
\end{aligned}
\end{equation}

Solving for \( \tilde{\zeta}_k \) yields
\begin{align}
\label{eq.49}
\tilde{\zeta}_k^\star 
&= \frac{1}{\ln(2)} \Bigg[ 
\boldsymbol{w}_k^{\text{H}} \left( \boldsymbol{h}_k^{\text{H}} \tilde{y}_k \tilde{y}_k^* \boldsymbol{h}_k + \sum_{i \neq k} \boldsymbol{h}_k^{\text{H}} \tilde{y}_i \tilde{y}_i^{\text{H}} \boldsymbol{h}_k \right) \boldsymbol{w}_k \nonumber \\
&\quad + \sigma_k^2 \tilde{y}_k^* \tilde{y}_k 
- 2 \operatorname{Re} \left\{ \tilde{y}_k^* \boldsymbol{h}_k \boldsymbol{w}_k \right\}
\Bigg]^{-1} - 1.
\end{align}

Finally, aggregating the terms involving \( \boldsymbol{n} \) yields
\[
f_r(\boldsymbol{n}) = \sum_{k=1}^{K} \left( 2 \operatorname{Re} \left\{ \boldsymbol{n}^{\text{H}} (\tilde{\kappa}_k \boldsymbol{I} - \boldsymbol{D}_k) \boldsymbol{z} \right\} - \tilde{\kappa}_k \boldsymbol{n}^{\text{H}} \boldsymbol{n} \right).
\]

Let
\[
\boldsymbol{M} = \sum_{k=1}^{K} \tilde{\kappa}_k \boldsymbol{I}, \quad \boldsymbol{B} = \sum_{k=1}^{K} (\tilde{\kappa}_k \boldsymbol{I} - \boldsymbol{D}_k) \boldsymbol{z}.
\]

Then, the simplified objective for \( \boldsymbol{n} \) becomes
\[
f_r(\boldsymbol{n}) = 2 \operatorname{Re} \left\{ \boldsymbol{n}^{\text{H}} \boldsymbol{B} \right\} - \boldsymbol{n}^{\text{H}} \boldsymbol{M} \boldsymbol{n}.
\]

The derivative with respect to \( \boldsymbol{n}^{\text{H}} \) can be calculated as
\[
\frac{\partial f_r(\boldsymbol{n})}{\partial \boldsymbol{n}^{\text{H}}} = 2 \boldsymbol{B} - 2 \boldsymbol{M} \boldsymbol{n}.
\]

Setting the derivative to zero yields
\[
\boldsymbol{M} \boldsymbol{n} = \boldsymbol{B}.
\]

This trivially leads to
\begin{equation}
    \label{eq:n_update_1}
    \boldsymbol{n}^\star = \boldsymbol{M}^{-1} \boldsymbol{B} = \Bigg(\sum_{k=1}^{K} \tilde{\kappa}_k \boldsymbol{I}\Bigg)^{-1}  \sum_{k=1}^{K} (\tilde{\kappa}_k \boldsymbol{I} - \boldsymbol{D}_k) \boldsymbol{z}.
\end{equation}

This solution now ensures consistency across all \( K \) terms, yielding a unified \( \boldsymbol{n}^* \), for which the $t$-th iteration is given by
\begin{equation}
\label{eq.51}
\hat{\boldsymbol{n}}^{(t)} = \mathcal{P}_{\boldsymbol{n} \in \mathcal{N}}\left( \Bigg(\sum_{k=1}^{K} \tilde{\kappa}_k \boldsymbol{I}\Bigg)^{-1}  \sum_{k=1}^{K} (\tilde{\kappa}_k \boldsymbol{I} - \boldsymbol{D}_k) \boldsymbol{n}^{t-1}\right),
\end{equation}
where \(\mathcal{P}_{\boldsymbol{n} \in \mathcal{N}}\) denotes the projection onto the feasible set \(\mathcal{N}\).

The complete executable steps for the proposed approach are summarized in Algorithm \ref{alg:NQT}.

\begin{algorithm}[H]
\caption{Proposed Secure ISAC Algorithm}
\label{alg:NQT}
\begin{algorithmic}[1]
\STATE \textbf{Input:} Vector \(\boldsymbol{\mu}\)
\STATE Initialize $\boldsymbol{w}$ and $\boldsymbol{n}$ to a feasible value based on the related constraints 
\REPEAT
\STATE \textbf{Optimization Problem 1} ($\boldsymbol{n}$ is held fixed)
    \FOR{each $k$}
        \STATE Update auxiliary variable $\boldsymbol{z}_k$ according to eq. \eqref{eq.27.1}.
        \STATE Update auxiliary variable $y_k$ according to eq. \eqref{eq.29}.
        \STATE Update auxiliary variable $\tilde{\zeta}_k$ according to eq. \eqref{eq.30}.
        \STATE Update decision variable $\boldsymbol{w}_k$ according to eq. \eqref{eq.31}.
        \STATE Update gradient projection $\boldsymbol{w}_k$ from eq. \eqref{eq.32}.
        
    \ENDFOR

\STATE \textbf{Optimization Problem 2} ($\boldsymbol w_k$ is held fixed)
    \STATE Update auxiliary variable $\boldsymbol{z}_k$ according to eq. \eqref{eq.43.1}.
    \STATE Update auxiliary variable $\tilde{y}_k$ according to eq. \eqref{eq.45}.
    \STATE Update auxiliary variable $\tilde{\zeta}_k$ according to eq. \eqref{eq.49}.
    \STATE Update decision variable $\boldsymbol{n}$ according to eq. \eqref{eq:n_update_1}.

\UNTIL{the objective function in \eqref{eq.13} converges}
\end{algorithmic}
\end{algorithm}

\section{Performance Analysis}
\label{sec:performance_analysis}

To assess the performance of the proposed approaches, this section presents numerical results obtained via Monte Carlo simulations. 
These results are used to demonstrate the effectiveness of the developed beamforming and artificial noise design. 
Without loss of generality, the entries of the channel matrix \(\boldsymbol{h}_k\) are modeled as independent and identically distributed (i.i.d.) complex Gaussian random variables following $\boldsymbol{h}_k \sim \mathcal{CN}(0,1)$. 
The \ac{ISAC} base station is assumed to employ a uniform linear array (ULA) with half-wavelength spacing between antenna elements. For simplicity, the noise variance at the targets is assumed to match that of the intended \acp{CU}. 
The parameters used for the analysis are summarized below in Table \ref{tab:simulation_parameters}.

\begin{figure}[H]
\centering
\includegraphics[width=\columnwidth]{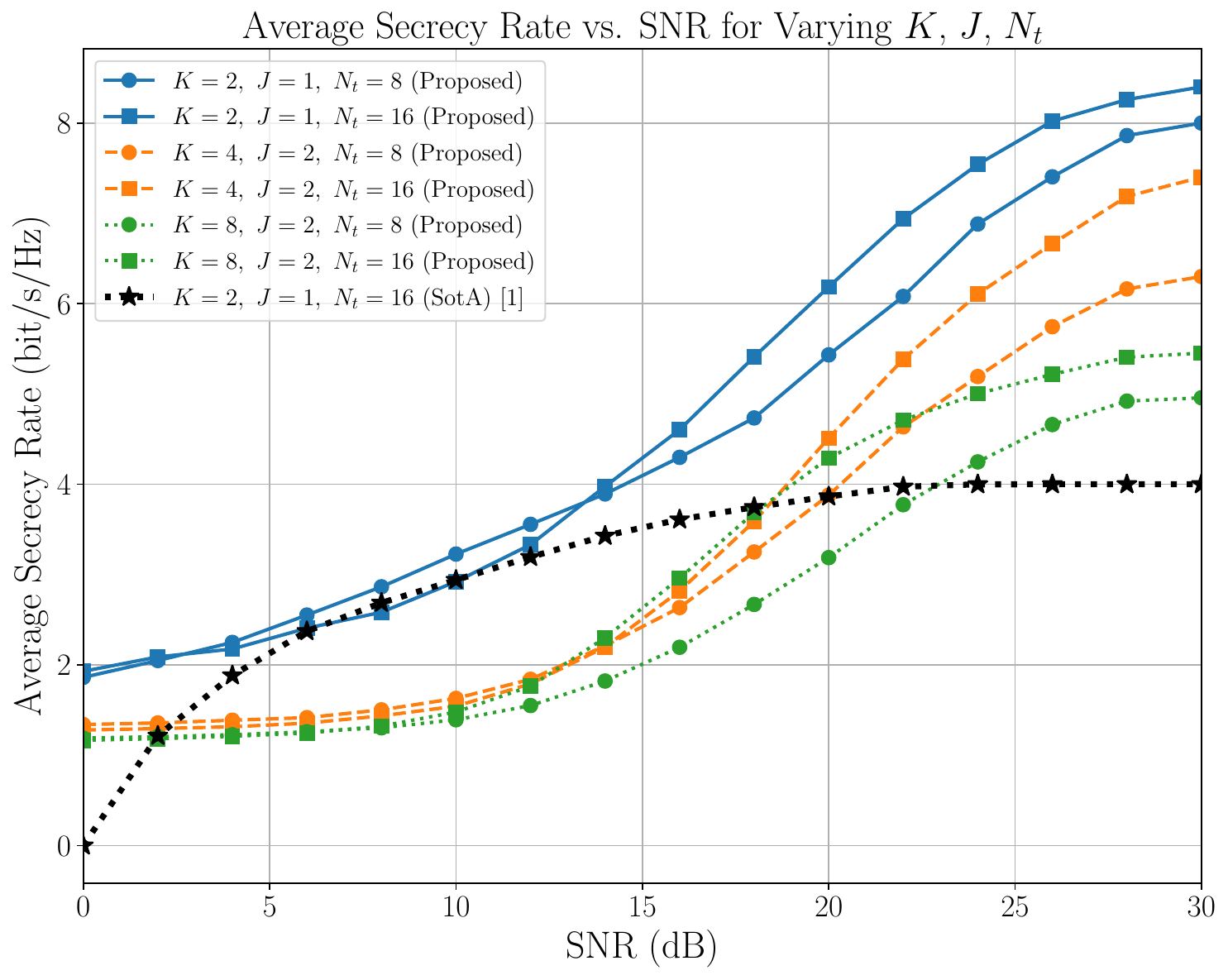}
\vspace{-4ex}
\caption{Average Secrecy Rate performance of the proposed Algorithm \ref{alg:NQT} under varying numbers of users \(K\), targets $J$, and transmit antennas $N_t$, compared with a representative \ac{SotA} method.}
\label{fig:secrecyRate}
\vspace{1ex}
\includegraphics[width=\columnwidth]{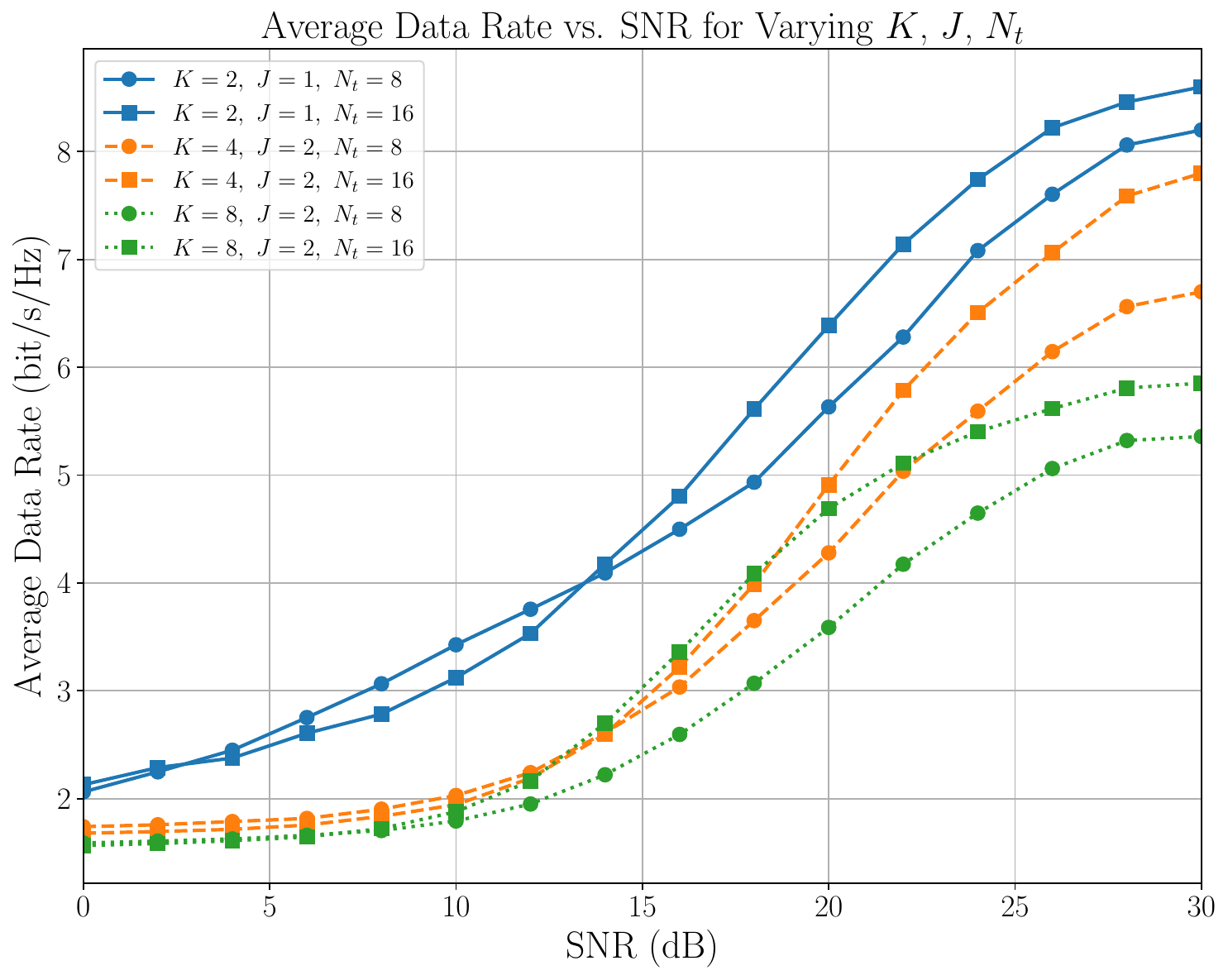}
\vspace{-4ex}
\caption{Average Data Rate performance of the proposed Algorithm \ref{alg:NQT} under varying numbers of users \(K\), targets $J$, and transmit antennas $N_t$.}
\label{fig:dataRate}
\vspace{1ex}
\includegraphics[width=\columnwidth]{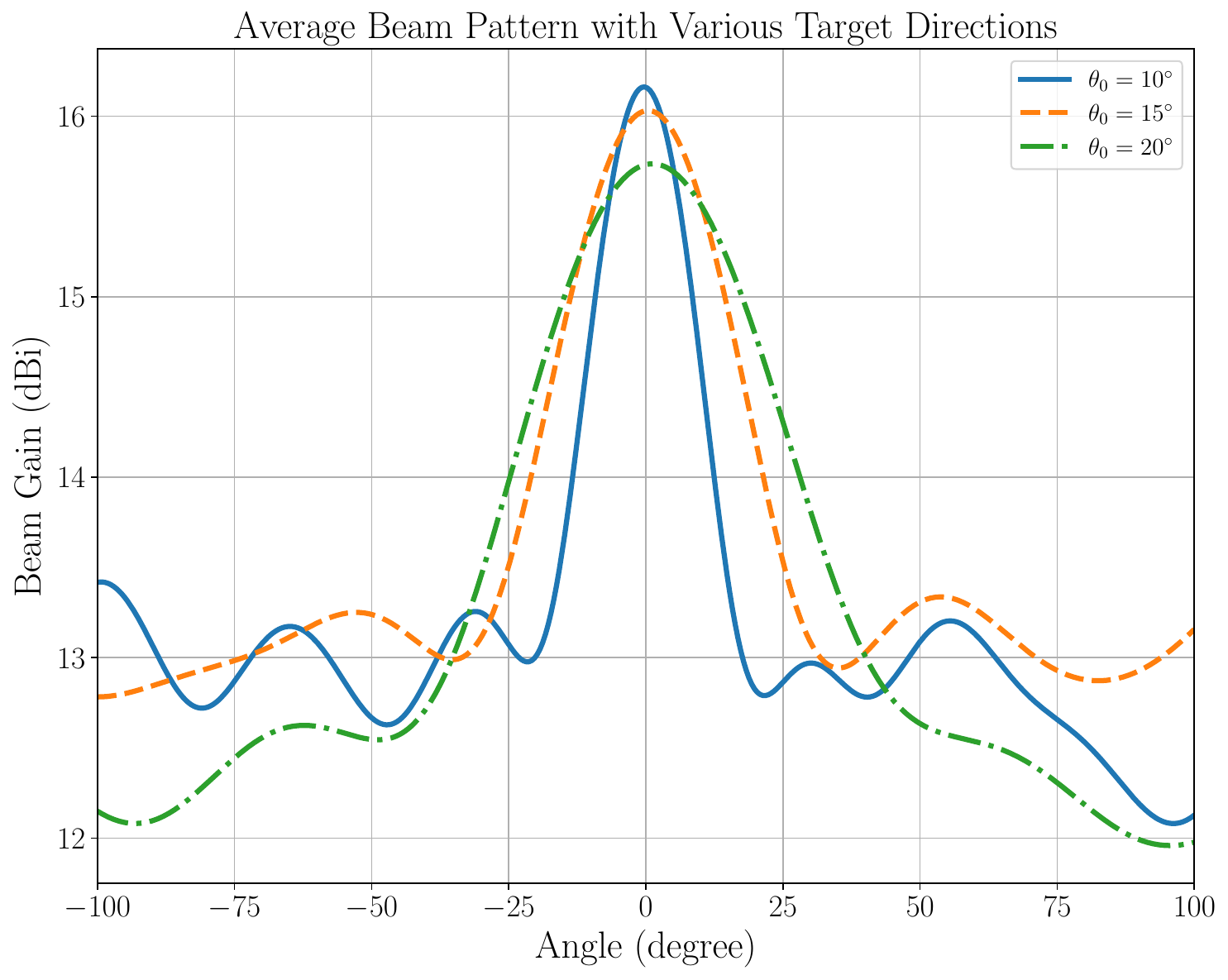}
\vspace{-4ex}
\caption{Beam Gain of the proposed Algorithm \ref{alg:NQT} for various target directions with $N_t = 16$, $K = 4$ and $J = 1$.}
\label{fig:beamGain}
\end{figure}

First, Fig. \ref{fig:secrecyRate} presents results for the average secrecy rate for a varying number \acp{UE}, eavesdroppers, and transmit antennas.
As highlighted in the aforementioned figure, the average secrecy rate increases when the number of transmit antennas increases due to the larger number of \ac{DoF} available in the system, while having more eavesdroppers and more users decreases the secrecy rate as a whole.
In addition, the proposed method outperforms the \ac{SotA} work done in \cite{SuTWC2021} across all \acp{SNR} with a large gain of more than 10 dB, due to the flooring of the \ac{SotA} technique, an average secrecy rate of around 4 bit/s/Hz.

Next, Fig. \ref{fig:dataRate} presents the performance of the proposed Algorithm \ref{alg:NQT} in terms of the average data rate as a metric for evaluating the communications performance of the system.

A similar trend to the average secrecy rate in Fig. \ref{fig:secrecyRate} is seen here as the average data rate increases when the number of transmit antennas increase due to the larger number of \ac{DoF} available in the system, while having more eavesdroppers and more users decreases the secrecy rate as a whole.
In addition, the average data rate is almost identical to the average secrecy rate, which means that very little data is leaked to eavesdroppers according to equation \eqref{eq.secrecy_rate}, with the users achieving almost a full rate within the system.

Figure \ref{fig:beamGain} illustrates the sensing performance by showing the beam gain across various eavesdropper directions of interest. The SINR threshold for each legitimate user is fixed at 20 dB. For reference, a narrow beam pattern, assuming perfect knowledge of the target direction at the base station, is used as a benchmark.
\vspace{-2ex}
\begin{table}[H]
\centering
\caption{Simulation Parameters for Joint Optimization}
\vspace{-2ex}
\label{tab:simulation_parameters}
\begin{tabular}{|c|c|c|}
\hline
\textbf{Parameter} & \textbf{Symbol} & \textbf{Value} \\
\hline
Number of users & $K$ & 2, 4, 8 \\
Number of eavesdroppers & $J$ & 1, 2 \\
Transmit antennas & $N_T$ & 8, 16, 18 \\
Total available power & $P_A$ & 20 dBm \\
Direction of eavesdroppers & $\theta_j$ & $[-90^\circ, +90^\circ]$ \\
Mainlobe leakage range & $\theta_0$ & $[0^\circ, 20^\circ]$ \\
Secrecy SINR threshold & $\beta_j$ & 0.1  \\
Minimum sensing gain & $\eta_j$ & 2 \\
Path loss (eavesdropper) & $\alpha_j$ & 1 \\
Fairness constraint threshold & $\xi_F$ & 0.5 \\
Entropy regularization weight & $\nu$ & 0.01 \\
\hline
\end{tabular}
\end{table}

\begin{figure}[H]
\centering
\includegraphics[width=\columnwidth]{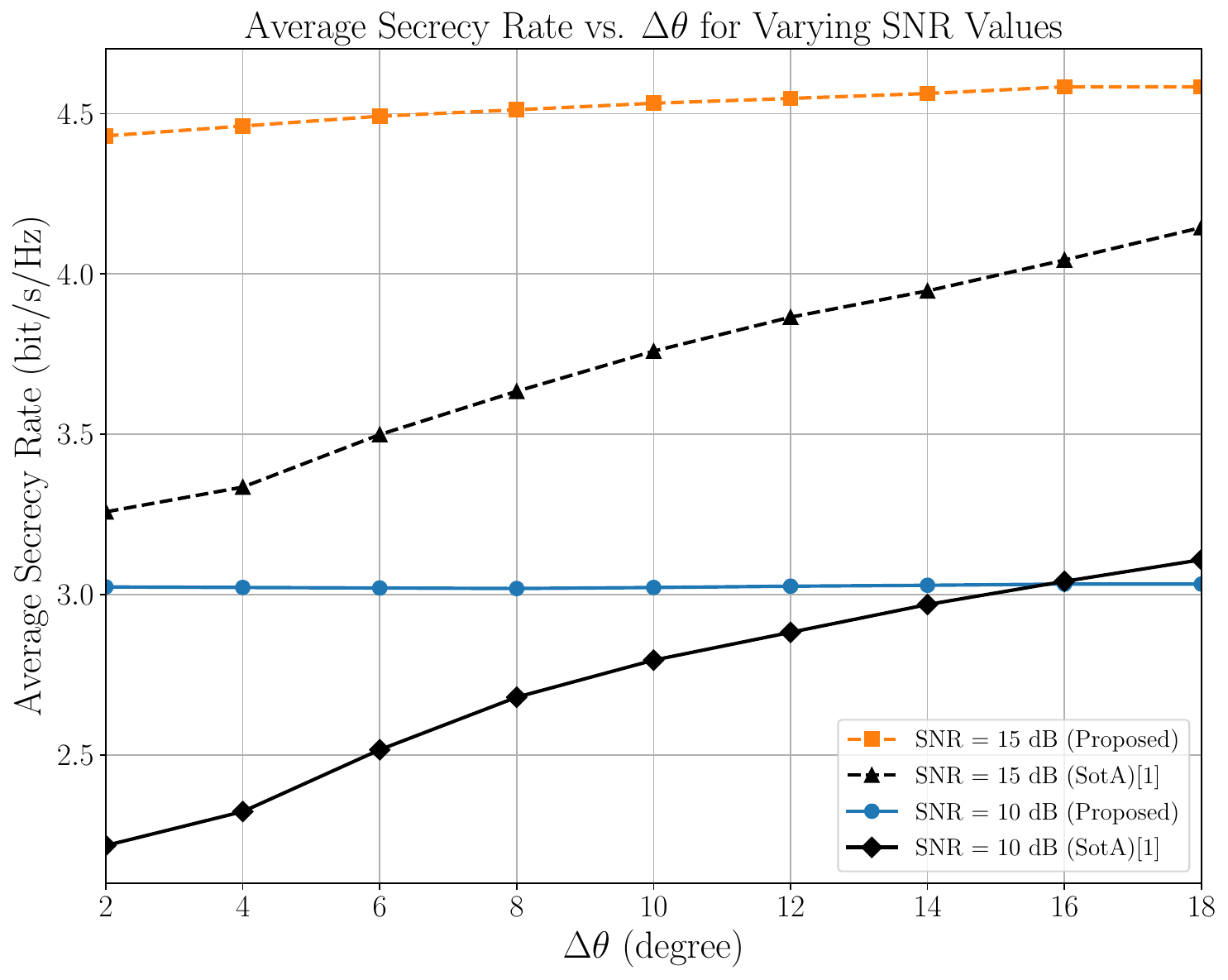}
\vspace{-5ex}
\caption{Average Secrecy Rate performance vs. varying angles of the proposed Algorithm \ref{alg:NQT} with $N_t = 18$, $K = 4$ and $J = 1$.}
\label{fig:deltaTheta}
\vspace{1ex}
\includegraphics[width=\columnwidth]{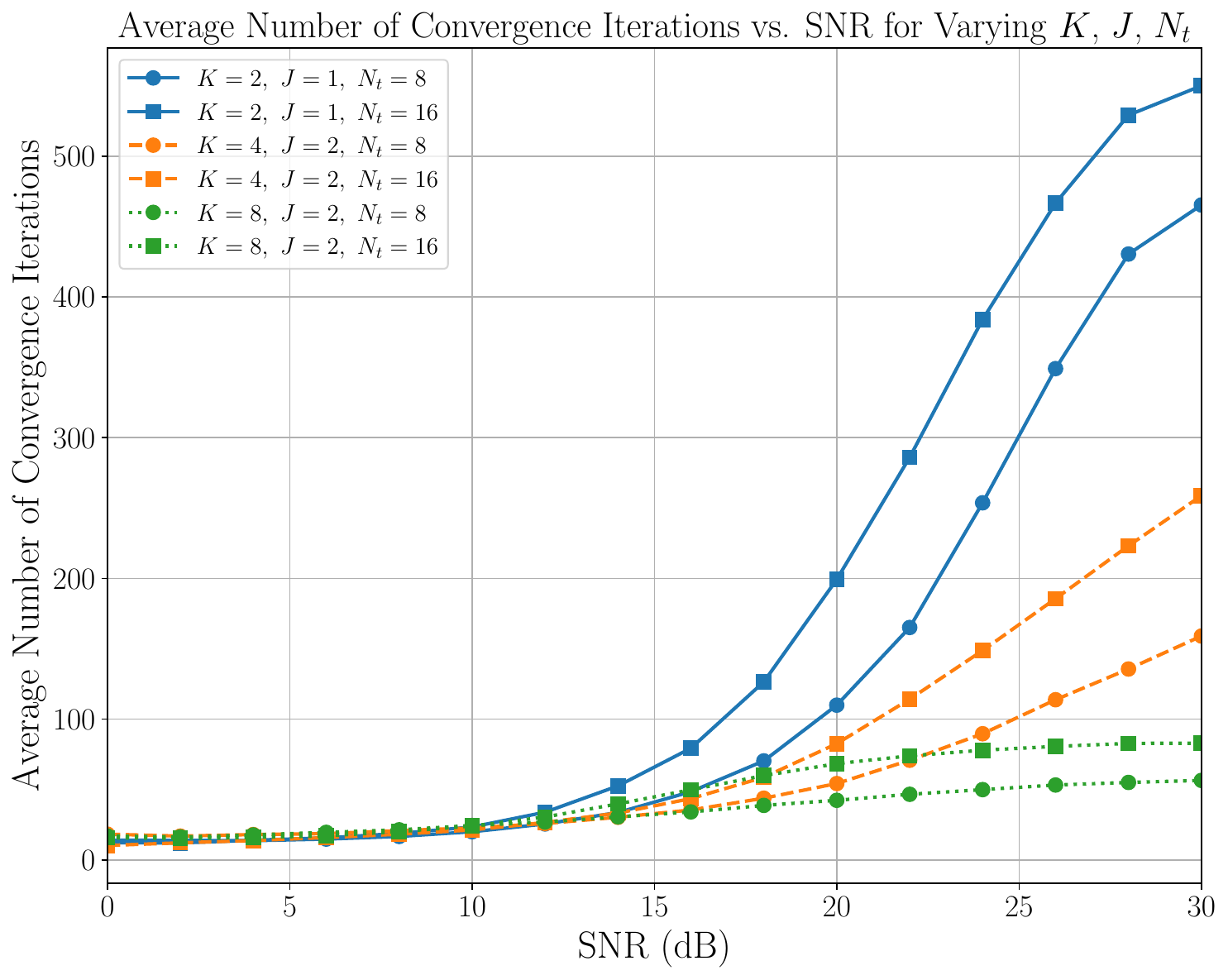}
\vspace{-5ex}
\caption{Convergence behavior of the proposed Algorithm \ref{alg:NQT} under a varying $K$, $J$ and $N_t$.}
\label{fig:convergence}
\end{figure}
\vspace{-2ex}

In contrast, the proposed algorithms produce wide main-lobe beam patterns that deliver consistent power levels across the entire angular range from $-90^\circ$ to $+90^\circ$, enabling robust sensing. While some reduction in beam gain is observed as the region of interest widens, the results confirm that high gains are preserved even for $\theta_0 = 20^\circ$, indicating reliable detection performance over a broad angular span.

Additionally, Fig.\ref{fig:deltaTheta} illustrates how the average secrecy rate varies with respect to changes in $\Delta \theta$. As evident from the figure, the proposed algorithm consistently outperforms the \ac{SotA} across all values of $\Delta \theta$, achieving a higher average secrecy rate throughout. Notably, thanks to the integration of artificial noise, the proposed method exhibits a more stable performance as $\Delta \theta$ increases, in contrast to the significant fluctuations observed in the \ac{SotA}.

Finally, Fig. \ref{fig:convergence} presents the convergence behavior of the proposed algorithm.
Due to the alternating nature of the proposed algorithm, the algorithm converges much faster for larger systems due to the larger number of \acp{DoF} in the solution space, while smaller systems require a much larger number of iterations to converge.

\section{Conclusion}
\label{sec:conclusion}

This work presented a novel secure MU-ISAC framework that jointly maximizes secrecy rate under communication and sensing constraints while promoting user fairness via an entropy-regularized Jain’s index. To solve the resulting non-convex problem, we proposed an efficient dual fractional programming method based on a non-homogeneous quadratic transform, enabling closed-form, scalable optimization of secure beamforming and artificial noise. We further introduced an iterative fairness–throughput tradeoff scheme to analyze system behavior across operating regimes. 
Simulation results corroborate the effectiveness of the proposed approach, demonstrating substantial performance gains in terms of average secrecy rate, data throughput, and beamforming accuracy, thus highlighting its suitability for secure and fair \ac{ISAC} system design.

\section*{Acknowledgment}
This work was funded by the German Federal Ministry of Education and Research (grant 16KISK231), the German Research Foundation (Germany's Excellence Strategy–EXC2050/1–ProjectID 390696704–Cluster of Excellence CeTI of Dresden, University of Technology), and based on the budget passed by the Saxon State Parliament.

\bibliographystyle{IEEEtran}
\bibliography{references}

\end{document}